# Reconstruction of Solar Extreme Ultraviolet Flux 1740–2015

Leif Svalgaard[1] (leif@leif.org)

[1] Stanford University, Cypress Hall C13, W.W. Hansen Experimental Physics Laboratory, Stanford, CA 94305, USA

**Abstract:**

Solar Extreme Ultraviolet (EUV) radiation creates the conducting E–layer of the ionosphere, mainly by photo ionization of molecular Oxygen. Solar heating of the ionosphere creates thermal winds which by dynamo action induce an electric field driving an electric current having a magnetic effect observable on the ground, as was discovered by G. Graham in 1722. The current rises and sets with the Sun and thus causes a readily observable diurnal variation of the geomagnetic field, allowing us the deduce the conductivity and thus the EUV flux as far back as reliable magnetic data reach. High–quality data go back to the 'Magnetic Crusade' of the 1830s and less reliable, but still usable, data are available for portions of the hundred years before that. J.R. Wolf and, independently, J.–A. Gautier discovered the dependence of the diurnal variation on solar activity, and today we understand and can invert that relationship to construct a reliable record of the EUV flux from the geomagnetic record. We compare that to the F10.7 flux and the sunspot number, and find that the reconstructed EUV flux reproduces the F10.7 flux with great accuracy. On the other hand, it appears that the Relative Sunspot Number as currently defined is beginning to no longer be a faithful representation of solar magnetic activity, at least as measured by the EUV and related indices. The reconstruction suggests that the EUV flux reaches the same low (but non–zero) value at every sunspot minimum (possibly including Grand Minima), representing an invariant 'solar magnetic ground state'.

# 1. Introduction

Graham (1724) discovered that the Declination, i.e. the angle between the horizontal component of the geomagnetic field (as shown by a compass needle) and true north, varied through the day. Canton (1759) showed that the range of the daily variation varied with the season, being largest in summer. Lamont (1851) noted that the range had a clear ~10–year variation, whose amplitude Wolf (1852a, 1857) and Gautier (1852) found to follow the number of sunspots varying in a cyclic manner discovered by Schwabe (1844). Thus was found a relationship between the diurnal variation and the sunspots "not only in average period, but also in deviations and irregularities" establishing a firm link between solar and terrestrial phenomena and opening up a whole new field of science. This was realized immediately by both Wolf and Gautier and recognized by many distinguished scientists of the day. Faraday wrote to Wolf on 27th August, 1852 (Wolf, 1852b):

> *I am greatly obliged and delighted by your kindness in speaking to me of your most remarkable enquiry, regarding the relation existing between the condition of the Sun and the condition of the Earths magnetism. The discovery of periods and the observation of their accordance in different parts of the great system, of which we make a portion, seem to be one of the most promising methods of touching the great subject of terrestrial magnetism...*

Wolf soon found (Wolf, 1859) that there was a simple, linear relationship between the yearly average amplitude, $v$, of the diurnal variation of the Declination and his relative sunspot number, $R$: $v = a + bR$ with coefficients $a$ and $b$, allowing him to calculate the terrestrial response from his sunspot number, determining $a$ and $b$ by least squares. He marveled "Who would have thought just a few years ago about the possibility of computing a terrestrial phenomenon from observations of sunspots".

Later researchers, (e.g. Chree, 1913; Chapman *et al.*, 1971), wrote the relationship in the equivalent form $v = a(1 + mR/10^4)$ separating out the solar modulation in the unit–independent parameter $m$ (avoiding decimals using the device of multiplying by $10^4$) with, it was hoped, local influences being parameterized by the coefficient $a$. Chree also established that $a$ and $m$ for a given station (geomagnetic observatory) were the same on geomagnetically quiet and geomagnetic disturbed days, showing that another relationship found with magnetic disturbances (Sabine, 1852) hinted at a different nature of *that* solar–terrestrial relation; a difference that for a long time was not understood and that complicates analysis of the older data (Macmillan and Droujinina, 2007).

Stewart (1882) suggested that the diurnal variation was due to the magnetic effect of electric currents flowing in the high atmosphere, such currents arising from electromotive forces generated by periodic (daily) movements of an electrically conducting layer across the Earth's permanent magnetic field. The next step was taken independently by Kennelly (1902) and Heaviside (1902) who pointed out that if the upper atmosphere was electrically conducting it could guide radio waves round the curvature of the Earth thus explaining the successful radio communication between England and Newfoundland established by Marconi in 1901. It would take another three decades before the notion of conducting ionospheric layers was clearly understood and accepted (Appleton, Nobel Lecture, 1947): the E–layer electron density and conductivity start to increase at sunrise, reach a maximum near noon, and then wane as the Sun sets; the variation of the conductivity through the sunspot cycle being of the magnitude required to account for the change with the sunspot number of the magnetic effects measured on the ground.

The Solar Extreme Ultraviolet (EUV) radiation causes the observed variation of the geomagnetic field at the surface through a complex chain of physical connections (as first suggested by Schuster (1908)), see Figure 1. The physics of most of the links of the chain is reasonably well–understood in quantitative detail and can often be successfully modeled. We shall use this chain *in reverse* to deduce the EUV flux from the geomagnetic variations, touching upon several interdisciplinary subjects.

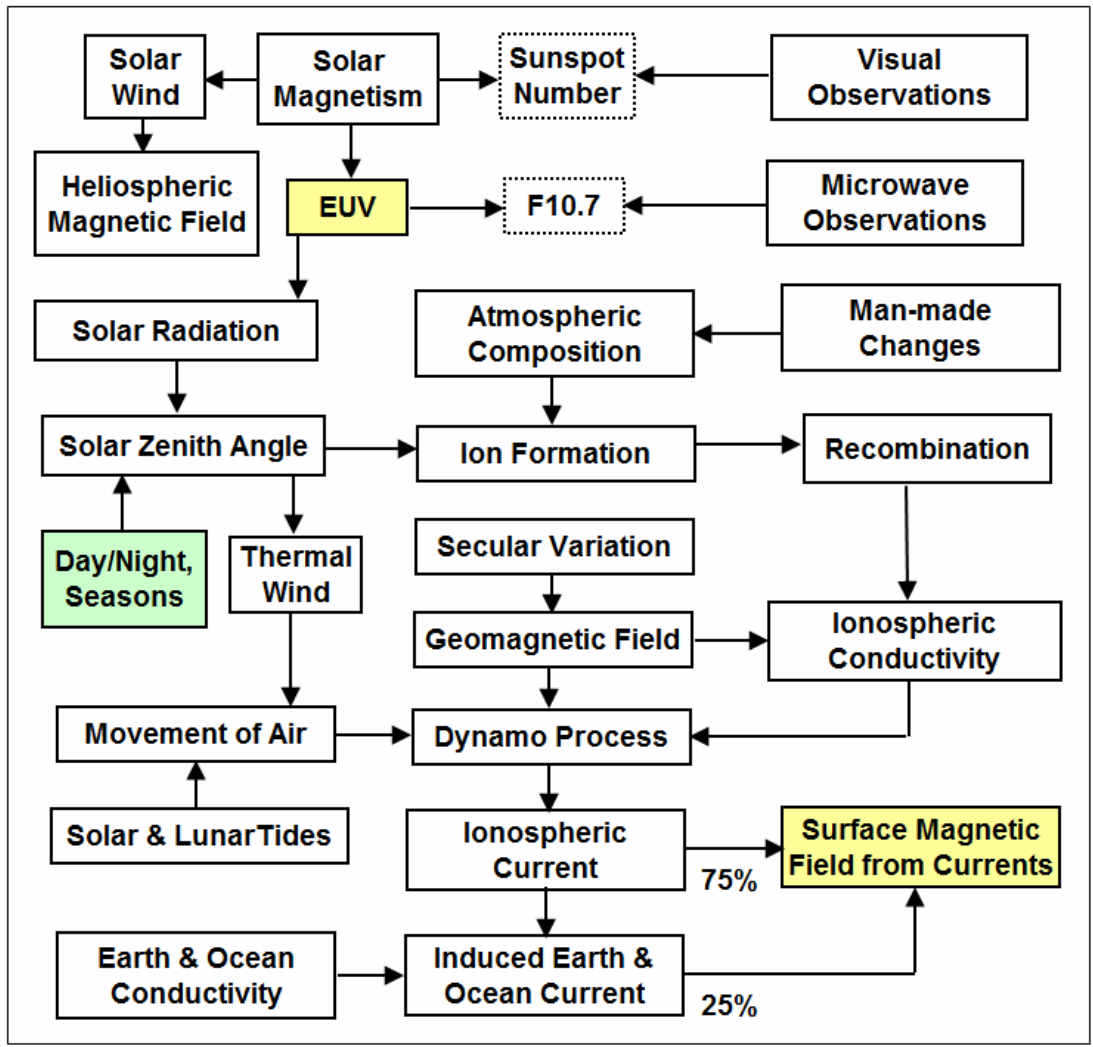

Figure 1: Block diagram of the entities and processes causally connecting variation of the solar magnetic field to the regular diurnal variation of the geomagnetic field. The effective ionospheric conductivity is a balance between ion formation and recombination. The movement of electrons across the geomagnetic field drives an efficient dynamo providing the electromotive force for the ionospheric currents giving rise to the observed diurnal variations of the geomagnetic field. The various blocks are further described in the text.

## 2. The Ionospheric E–Layer

The dynamo process takes place in the dayside E–layer where the density, both of the neutral atmosphere and of electrons is high enough. The conductivity at a given height is roughly proportional to the electron number density $N_e$. In the dynamo region (at 105 km altitude), the dominant plasma species is molecular oxygen ions, $O_2^+$, produced by photo ionization (by photons of wavelength $\lambda$ of 102.7 nm or less (Samson and Gardner, 1975)) at a rate $J$ per unit time $O_2 + h\nu \xrightarrow{J} O_2^+ + e^-$ and lost through recombination with electrons at a rate $\alpha$ per unit time $O_2^+ + e^- \xrightarrow{\alpha} O + O$, in the process producing the Airglow. The rate of change of the number of ions, $N_i$, $dN_i/dt$ and of electrons, $N_e$, $dN_e/dt$ are given by $dN_i/dt = J\cos(\chi) - \alpha N_i N_e$ and $dN_e/dt = J\cos(\chi) - \alpha N_e N_i$, respectively, where we have ignored motions into or out of the layer. Since the Zenith angle $\chi$ changes but slowly, we have a quasi steady–state (with a time constant of order $1/(2\alpha N) \approx 1$ minute), in which there is no net electric charge, so $N_i = N_e = N$. In steady state $dN/dt = 0$, so that the equations can both be written $0 = J\cos(\chi) - \alpha N^2$, or when solving for the number of electrons $N = \sqrt{J\cos(\chi)/\alpha}$ (using the sufficient approximation of a flat Earth with a layer of uniform density). Since the conductivity, $\Sigma$, depends on the number of electrons we expect that $\Sigma$ should scale with the square root $\sqrt{J}$ of the overhead EUV flux (Yamazaki and Kosch, 2014). Even if the exponent is not quite ½ (*e.g.* Ieda *et al.*, 2014), that is not critical to and has no influence on the result of our analysis.

The magnitude, $A$, of the variation of the East Component due to the dynamo process is given by $A = \mu_o \Sigma U B_z$ (Takeda, 2013) where $\mu_o$ is the permeability of the vacuum ($4\pi \times 10^7$), $\Sigma$ is the height–integrated effective ionospheric conductivity (in S), $U$ is the zonal neutral wind speed (m/s), and $B_z$ is the vertical geomagnetic field strength (nT). The conductivity is a highly anisotropic tensor and in the E–layer the electrons begin to gyrate and drift perpendicular to the electric field, while the ions still move in direction of the electric field; the difference in direction is the basis for the Hall conductivity $\Sigma_H$, which is there larger than the Pedersen conductivity $\Sigma_P$. The combined conductivity then becomes $\Sigma = \Sigma_P + \Sigma_H^2/\Sigma_P$ (Koyama *et al.*, 2014; Ieda *et al.*, 2014; Takeda, 2013; Maeda, 1977).

The various conductivities depend on the ratio between the electron density $N$ and the geomagnetic field $B$ times a slowly varying dimensionless function involving ratios of gyro frequencies $\omega$ and collision frequencies $\nu$: $N/B \times f(\omega_e, \nu_{en\perp}, \omega_i, \nu_{in})$ (Richmond, 1995) such that, to first approximation, $\Sigma \sim N/B$ (Clilverd *et al.*, 1998), with the result that the magnitude $A$ only depends on the electron density and the zonal neutral wind speed. On the other hand, simulations by Cnossen *et al.* (2012) indicate a stronger dependence on $B$ (actually on the nearly equivalent magnetic dipole moment of the geomagnetic field $M$), $\Sigma \propto M^{-1.5}$, leading to a dependence of $A$ on $M$: $A \propto M^{-0.85}$, and thus expected to cause a small secular increase of $A$ as $M$ is decreasing over time This stronger dependence is barely, if at all, seen in the data. We return to this point in Section 7. The purported near–cancellation of $B$ is not perfect, though, depending on the precise geometry of the field. In addition, the ratio between internal and external current intensity varies with location. The net result is that $A$ can and does vary somewhat from location to location even for given $N$ and $U$. Thus a normalization of the response to a reference location is necessary, as discussed in detail in Section 7.

## 3. The EUV Emission Flux

The Solar EUV Monitor (CELIAS/SEM) onboard the SOHO spacecraft at Lagrange Point L1 has measured the integrated solar EUV emission in the 0.1–50 nm band since 1996 (Judge *et al.*, 1998). The calibrated flux at a constant solar distance of 1 AU can be downloaded from http://www.usc.edu/dept/space_science/semdatafolder/long/daily_avg/. For our purpose, we reduce all flux values to the Earth's distance. The main degradation of the SEM sensitivity is attributed to build–up (and subsequent polymerization by UV photons) of a hydrocarbon contaminant layer on the entrance filter, and is mostly corrected for using a model of the contaminant deposition. We estimate any *residual* degradation by monitoring the ratio between the reported SEM EUV flux (turquoise curve in Figure 2) and the F10.7 microwave flux, Figure 2 (purple points), and adjusting the SEM flux accordingly (red curve). The issue of degradation of SEM has been controversial (Lean *et al.*, 2011; Emmert *et al.*, 2014; Didkovsky and Wieman, 2014) and is, perhaps, still not completely resolved (Wieman *et al.*, 2014). We constrain the SEM flux to match F10.7 as suggested by Emmert *et al.* (2014).

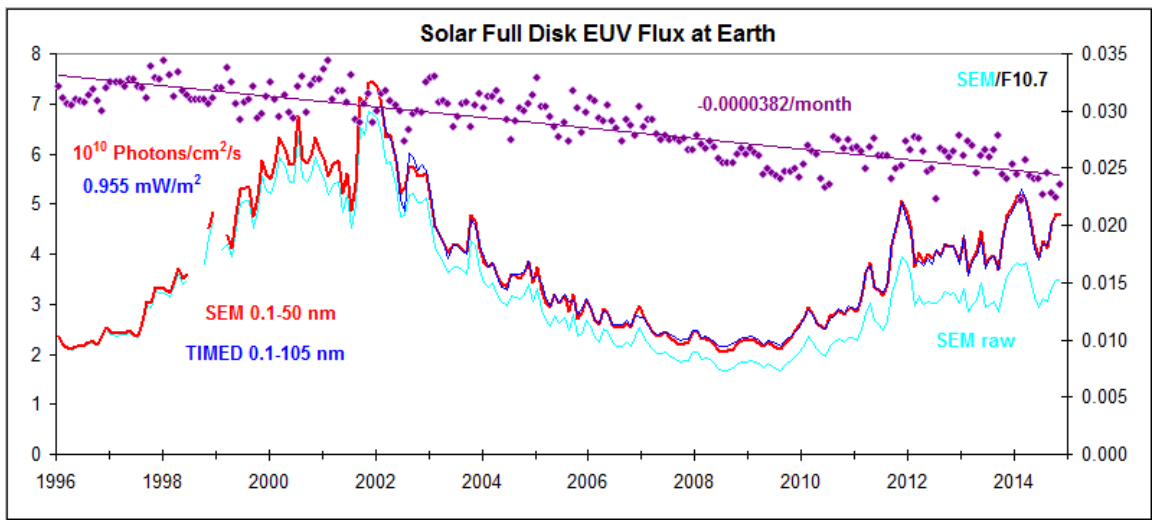

Figure 2: Integrated CELIAS–SEM absolute solar EUV flux in the 0.1–50 nm band (turquoise curve) uncorrected for *residual* degradation of the instrument and the corrected flux (red curve) as derived from the decrease of the ratio between the raw EUV flux and the F10.7 microwave flux (purple points). The degradation–corrected integrated flux in the 0.1–105 nm band measured by TIMED (blue curve) matches the corrected SEM flux requiring only a simple, constant scaling factor. All data are as measured at Earth rather than at 1 AU.

The Solar EUV Experiment (SEE) data from the NASA 'Thermosphere, Ionosphere, Mesosphere Energetics and Dynamics (TIMED)' mission (Woods *et al.*, 2005) provide, since 2002, daily averaged solar irradiance in the 0.1–105 nm band with corrections applied for degradation and atmospheric absorption (with flare spikes removed) and can be downloaded from http://lasp.colorado.edu/home/see/data/. The SEE flux (Figure 2, blue curve) is very strongly linearly correlated with (and simply proportional to) our residual–degradation–corrected SEM flux (with coefficient of determination $R^2 = 0.99$) and thus serves as validation of the corrected SEM data. SEM data is in units of photons/cm$^2$/s while SEE data is in units of mW/m$^2$. Using a reference spectrum, the two scales can be converted to each other ($10^{10}$ photons/cm$^2$/s ↔ 0.955 mW/m$^2$). We shall here use a composite of the SEM and SEE data in SEM units. All data used are supplied in the Supplementary Data Section of this paper.

## 4. The F10.7 Flux Density

The λ10.7 cm microwave flux (F10.7) has been routinely measured in Canada (first at Ottawa and then at Penticton) since 1947 and is an excellent indicator of the amount of magnetic activity on the Sun (Tapping, 1987, 2013). The 10.7 cm wavelength corresponds to the frequency 2800 MHz. Measurements of the microwave flux at several frequencies from 1000 MHz (λ30 cm) to 9400 MHz (λ3.2 cm), straddling 2800 MHz, have been carried out in Japan (first at Toyokawa and then at Nobeyama) since the 1950s (Shibasaki *et al.*, 1979) and allow a cross–calibration with the Canadian data. A 2% decrease of the 2800 MHz flux is indicated when the Canadian radiometer was moved from Ottawa to Penticton in mid–1991 (Svalgaard, 2010). We correct for this by reducing the Ottawa flux accordingly. As the morning and afternoon measurements at Penticton are, at times (especially during the snowy winter), afflicted with systematic errors (of unknown provenance) we only use the noon–values of the observed flux (not adjusted to a solar distance of 1 AU) and form a composite (updated through 2014) with the Japanese data (for 2000 and 3750 MHz) scaled to the Canadian 2800 MHz (Svalgaard, 2010; Svalgaard and Hudson, 2010), Figure 3, similar to the composite by Dudok de Wit *et al.* (2013).

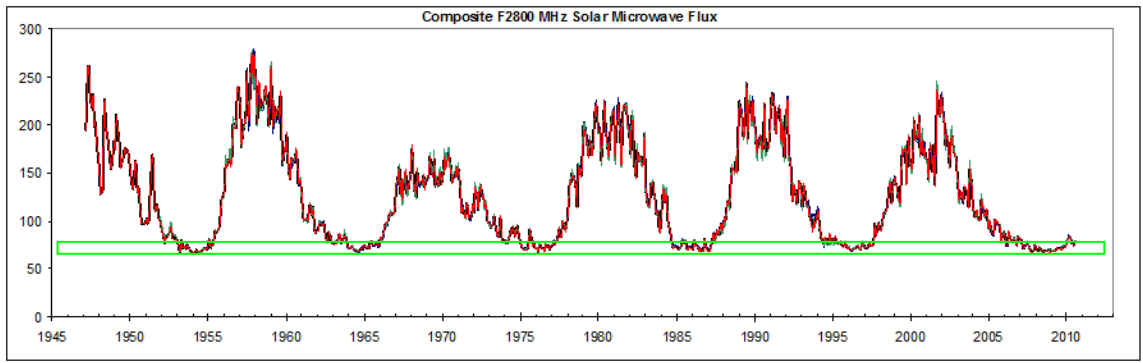

Figure 3: Composite 2800 MHz solar microwave flux (thin black curve) built from Canadian 2800 MHz flux (red curve), scaled Japanese 3750 MHz flux (green curve), and scaled Japanese 2000 MHz flux (blue curve), Svalgaard (2010). The match is so good that it is difficult to see the individual curves as they fall on top of each other. Note that the values at each sunspot minimum are very similar, without any long-term trend or inter-cycle variation.

The reported F10.7 data can be downloaded from the Dominion Radio Astrophysical Observatory at ftp://ftp.geolab.nrcan.gc.ca/data/solar_flux/daily_flux_values/. Although the absolute calibration of the observed flux density shows that the flux values must be multiplied by 0.9 (the 'URSI' adjustment), we follow tradition and do not apply this adjustment. The Japanese data can be downloaded from http://solar.nro.nao.ac.jp/norp/.

# 5. The $S_R$ Current System

More than 200 geomagnetic observatories around the world measure the variation of the Earth's magnetic field from which the *regular*, solar local time daily variations described by Canton (1759) and Mayaud (1965), $S_R$, can be derived. From the variation of the horizontal component $\Delta H$, one can derive the surface current density, $K$, for a corresponding equivalent thin–sheet electric current system overhead, $K$ (mA/m) = 1.59 $\Delta H$ (nT) = $2\Delta H/\mu_o$. This relationship is not unique; the current system is three–dimensional, and an infinite number of current configurations fit the magnetic variations observed at ground level. Measurements in space provide a much more realistic picture (Olsen, 1996) and the $S_R$ system is only a convenient *representation* of the true current.

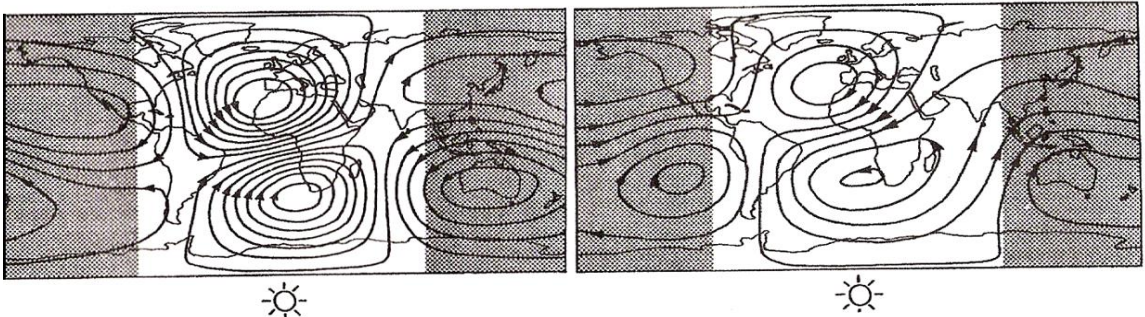

Figure 4: Streamlines of equivalent $S_R$ currents during equinox at 12 UT separately for the external primary (left) and the internal secondary (right) currents (Adapted after Malin, 1973).

Figure 4 (left) shows current streamlines of the equivalent $S_R$ current as seen from the Sun at (Greenwich) noon. This current configuration is fixed with respect to the Sun with the Earth rotating beneath it. The westward moving $S_R$ current vortex and the electrically conducting Earth interior (and ocean) act as a transformer with the E–layer as the primary winding and the conducting ground as the secondary winding, inducing electric currents at depth. The magnetic field of the secondary current (about 30% of that of the primary) adds to the magnetic field of the primary $S_R$ current. We are concerned only with the total variation resulting from superposition of the two components. In addition, we do not limit ourselves to the variation, $S_q$, of the so–called 'quiet days', as their level of quietness varies with time, but rather use data from all days when available (the difference is in any case small).

The $S_R$ current depends on season, i.e. the solar Zenith angle controlling the flux of EUV radiation onto the surface. The summer vortex is larger and stronger than the winter vortex and actually spills over into the winter hemisphere. The amplitude of the $S_R$ increases by a factor of two from solar minimum to solar maximum, mostly due to the solar cycle variation of conductivity caused by the solar cycle variation of the EUV flux (Lean *et al.*, 2003). In addition, the daytime vortices show a day–to–day variability, attributed to upward–traveling internal waves that are sensitive to varying conditions in the lower atmosphere.

Atmospheric magnetic tides (Love and Rigler, 2014) are global–scale waves excited by differential solar heating or by gravitational tidal forces of both the Moon and the Sun, and are sometimes studied using a wave–formalism on which we shall comment briefly:

The atmosphere behaves like a large (imperfect) waveguide closed at the surface at the bottom and open to space at the top, allowing an infinite number of atmospheric wave modes to be excited, but only low-order modes are important. There are two kinds of wave modes: class 1 waves (gravity waves), and class 2 waves (rotational waves); the latter owing their existence to the Coriolis Effect. Each mode is characterized by two numbers ($m$, $n$): a zonal wave number $n$ (positive for class 1 and negative for class 2 waves) and a meridional wave number $m$, with periods relative to one solar or lunar day, respectively. The fundamental solar diurnal tidal mode, known as $S_1$, that is most strongly excited is the (1, -2) mode, being an external mode of class 2 depending on solar local time. The largest solar semidiurnal wave, $S_2$, is a mode (2, 2) internal class 1 wave. The dominant migrating lunar tide, $M_2$, is a ~20 times smaller (2, 2) mode depending on lunar local time, and will not be considered further here.

## 6. The Diurnal Range of the Geomagnetic East Component

The $S_R$ current system rises with the Sun in the morning, with the pre–noon current at northern mid-latitudes running from north to south (in the opposite direction at southern latitudes) and when the Sun and the currents set, the afternoon current is from south to north (Figure 4). The magnetic effect due to these currents is at right angles to the current direction, i.e. east–west. Currents due to solar wind induced geomagnetic disturbances (Ring Current; electrojets) tend to flow east–west, so their magnetic effect is strongest in the north-south direction and generally lowest and rather disorganized in the east–west direction, hence have little effect on the average east–west magnetic variations. For this reason, the variation of the geomagnetic East–Component (and the almost equivalent Declination, Figure 5) is especially suited as a proxy for the strength of the $S_R$ current.

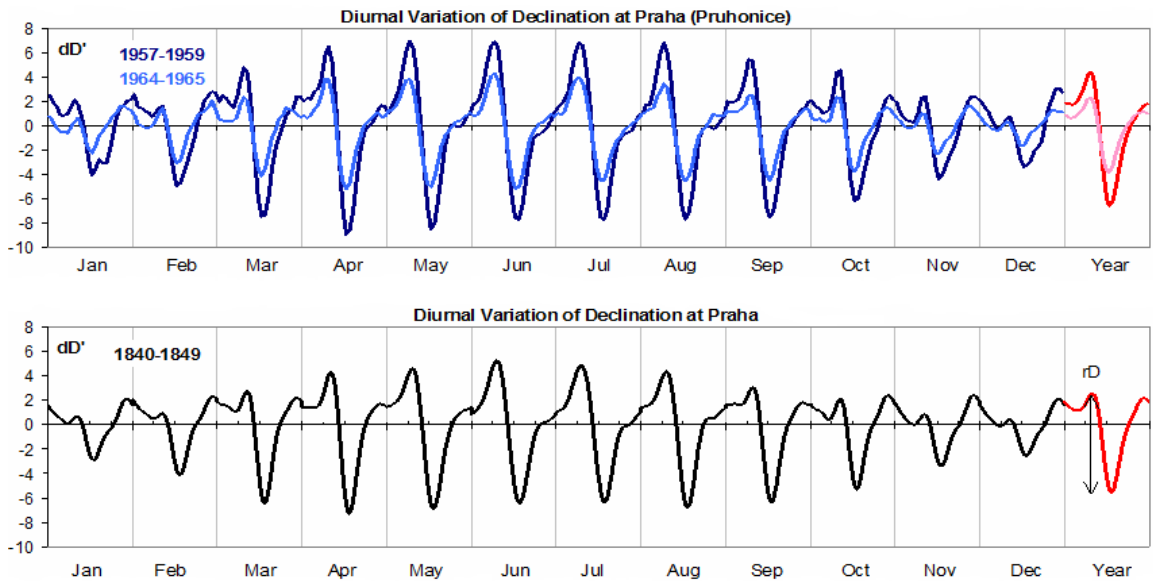

Figure 5: Diurnal variation of Declination (in arc minutes) at Prague per month. For each month is shown the variation (with respect to the daily mean) over one local solar day from midnight through noon to the following midnight. Top: modern data for low sunspot number (1964–1965, light blue) and for high sunspot number (1957–1959, dark blue). Bottom: average for the interval 1840–1849. The red

curves show the yearly averages. The range, *rD*, should be defined as the difference between the values of the pre–noon and post–noon local extrema, rather than simply between the highest and lowest values for day.

Some geomagnetic observatories report measurements of the Horizontal Component *H* and the Declination *D*, while others report the North and East Components, *X* and *Y*, determined by $X = H \cos(D)$, $Y = H \sin(D)$. For a small change $dD'$ (arc minutes) in *D*, the change in *Y* is often approximated by $dY = H \cos(D) \, dD'/3438'$. We convert all variations directly to force units (nT) without using the approximation, whenever possible. Many early observers did not measure *H*, but only *D*. We can still calculate *Y* because *H* can with sufficient accuracy as needed be determined for any location from historical spherical harmonics coefficients at any time in the past 400 years (Jackson *et al.*, 2000). Actually, there is a benefit to using the angle *D*, as angles do not need calibration. It is clear from Figure 5 that the measurements from the 1840s are accurate enough to show, even in minute detail, the same variations as the modern data and that the amplitude, and hence solar activity, back then was intermediate between that in 1964–1965 and 1957–1959.

In order to construct a long–term record we shall work with yearly averages of the range, *rY*, of the diurnal variation of the East–Component, defined as the unsigned difference between the values of the pre–noon and post–noon local extrema of *Y*. The values can be hourly averages or spot–values, the (small) difference corrected for by suitable normalization, if needed. Many older stations only observed a few times a day or twice, usually near the times of maximum excursions from the mean. As long as these observations were made at fixed times during the day, they can be used to construct a nominal daily range. Most long–running observatories had to be moved to replacement stations further and further away from their original locations due to electrical and urban disturbances, forming a station 'chain'. We usually normalize the data separately for replacement stations, except when they are co–located upgrades of the original station.

At mid-latitude stations (say around 35º latitude) the electric currents flow generally North-South over a relatively large range of latitudes, so most of the magnetic effect of the current will be in the East-West component. The magnetic effects of the auroral zone and equatorial ejectrojets as well as of the Ring Current are mostly in the North-South direction so are minimized by limiting our investigation to the East-component.

Another advantage of using the East component is that (at least generally before the 20[th] century) it often is based on observations of the Declination, which, being an angle does not require calibration (other than the trivial conversion from scale values) nor difficult temperature corrections. That said, the Declination *is* sensitive to the disturbing effects of nearby iron masses. A valid criticism of the use of the range in Declination, *rD*, is that it is the resultant of the pull of two force vectors: the (nearly constant) largely North-South horizontal force of the main geomagnetic field and the (varying during the day) largely East-West force of the magnetic effect of the $S_R$ current system, and that therefore the range of the angle in arc minutes varies with the horizontal force as well, in space and in time. François Arago wonderfully described (in the 1820s) how the range of the Declination he observed at the Paris Observatory increased by a factor of ten as the result of installation (later removed) of an iron stove in an adjoining room, the magnetic stove canceling out a large part of the natural horizontal force.

## 6.1. The Master Record

The German station chain (POT–SED–NGK) yields an almost unbroken data series extending over 125 years. The French station chain (PSM–VLJ–CLF) provides an even longer series, 130 years of high–quality data, (Fouassier and Chulliat, 2009). The diurnal variation of $\Delta Y$ (Figure 6) is essentially the same for both chains at both ends of the series. There is a clear 0.7 hour shift of CLF with respect to NGK due to modern daily records covering a UT–day rather than the local solar day. This has negligible impact on the range $rY$.

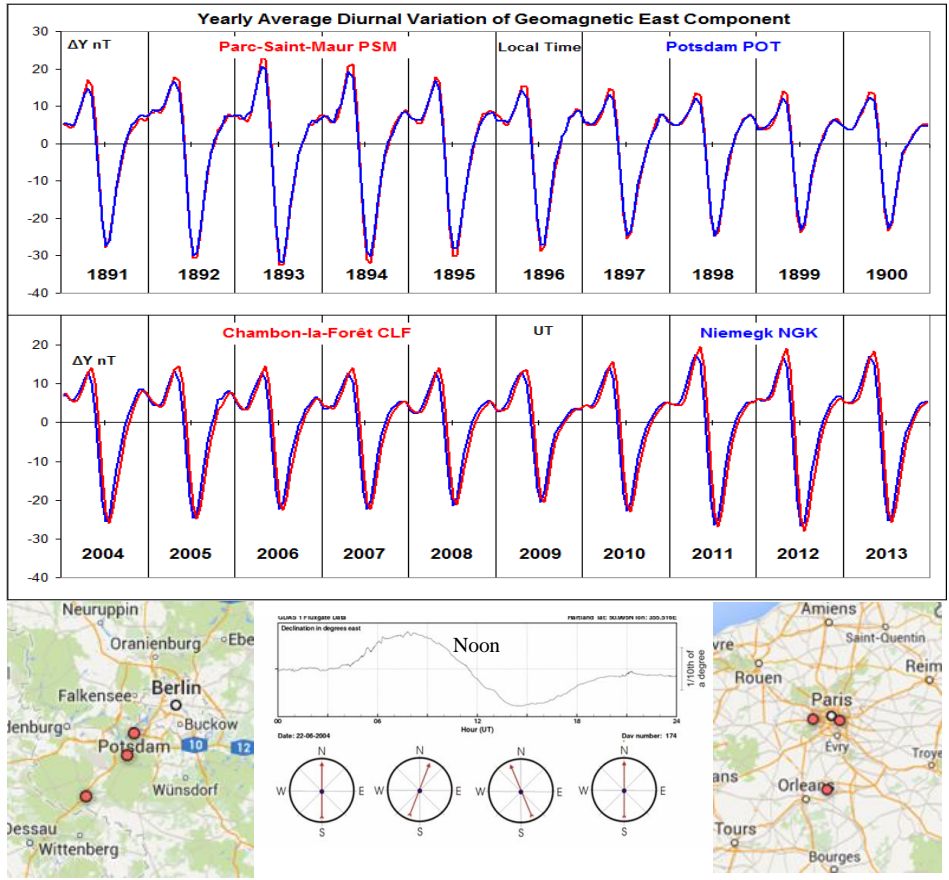

Figure 6: Yearly average diurnal variation, $\Delta Y$, of the East–Component for (top) observatories PSM (France) and POT (Germany) for each year of the decade 1891–1900, and for (center) observatories CLF (France) and NGK (Germany) for the decade 2004–2013. (Bottom) Observatory locations and schematic variation of the direction of the 'magnetic needle'.

The variation at the French stations is 5% larger than at the German stations. We form a simple composite Master–Record, adjusting the French stations down by 5%, Figure 7. The Master–Record is thus fundamentally and arbitrarily rooted in the German series. No further adjustments of the intra–chain records can be made (and none seems necessary), as the available data for individual stations in each chain do not overlap enough in time.

It is immediately apparent that there is very little, if any, variation of the range at sunspot minima (dashed line in lower panel of Figure 7). The lack of a trend in the mid-latitude geomagnetic response to solar activity in general has also been noted by Martini *et al.* (2015).

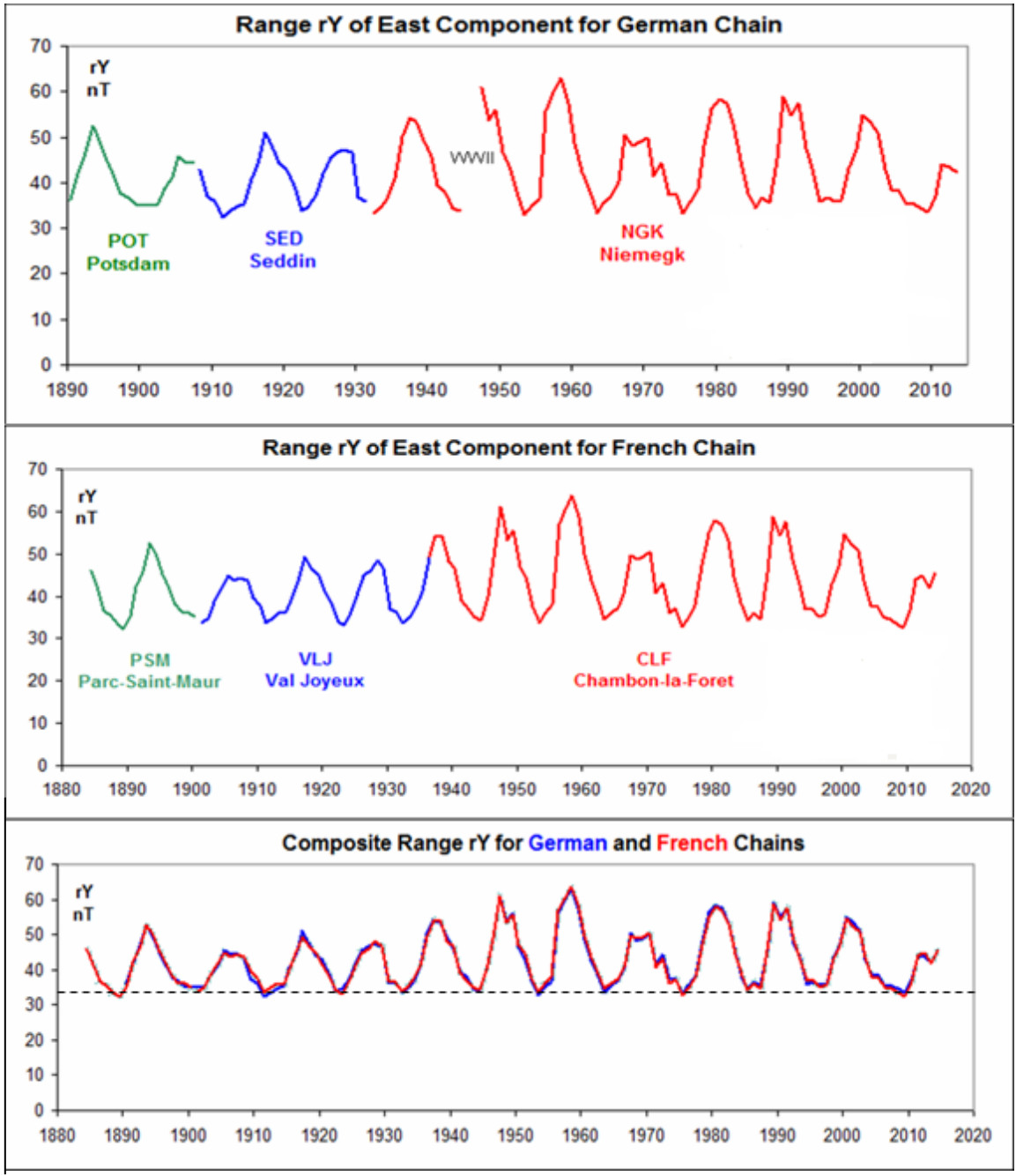

Figure 7: Yearly average diurnal range, *rY*, of the East–Component for German stations near Berlin (top panel), for French stations near Paris (middle panel), and a composite (bottom panel) being simply the average of the German records and of the (scaled down by 5%) French records.

## 7. Normalization of the Diurnal Range, *rY*

The composite shall serve as a *Master–Record* to which all other stations will be normalized. The vertical component in Central Europe over the time period of the Master–Record has increased by some 3%. We would expect a corresponding 2% decrease of the magnetic effect of the $S_R$ system over that time, or a 1.3 nT/century *decrease* that, however, does not seem to be visible in the data at sunspot minima. Other stations seem to show an *increase* of a similar amount (Macmillan and Droujinina, 2007; *Yamazaki and Kosch*, 2014) or no increase at all ("Sq(Y) did not increase significantly at observatories where the main field intensity decreased" (*Takeda*, 2013)). The issue is still open and several other variables could be in play, such as variation of the upper atmospheric wind patterns, changes in atmospheric composition, and changes in the altitude and/or density of the dynamo region (affecting the mix of Hall and Pedersen conductivities). Our position here shall be not to try to make ad–hoc corrections for the change of the main field.

The Eskdalemuir station (ESK) has been in almost continuous operation with good coverage since 1911. After getting correct data (MacMillan and Clarke, 2011), the normalization procedure begins with regression of the Master *rY* against the observatory (ESK in this case), Figure 8. Outliers, if any, are identified and omitted. We find that, almost always, the regression line goes through the origin within the uncertainty of the regression, so we force it through the origin (occasionally a better fit is a weak power law which we then use instead).

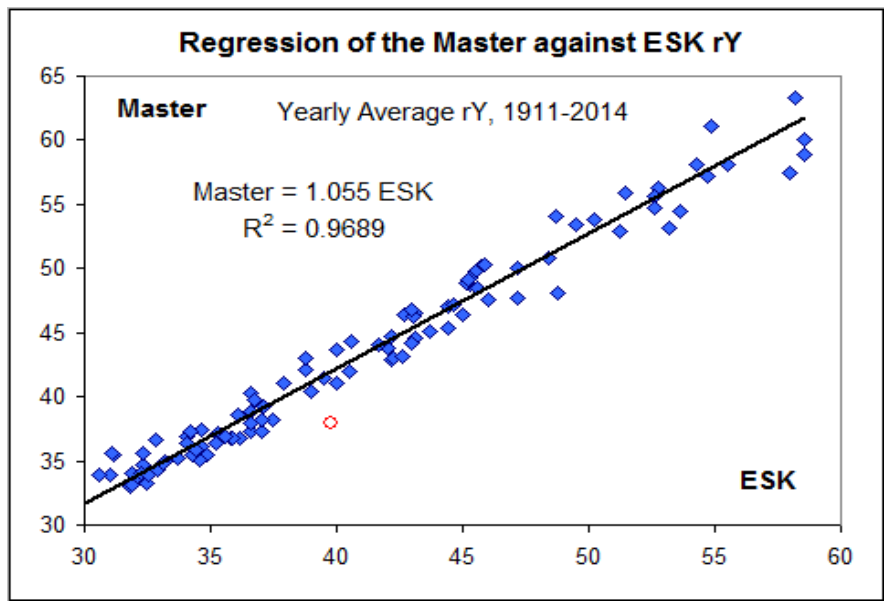

Figure 8: Linear regression of *rY* for the Master against Eskdalemuir (ESK). On account of significant missing data in 1984, the data point for that year (red circle) is a clear outlier and has not been included in the regression. The slope of the regression line indicates the factor by which to multiply the station value to normalize it to the Master Composite.

When the normalized data is plotted together with the Master Composite (e.g. Figure 9) a further visual quality control is performed and stations with large discrepancies (mostly of unknown causes) are omitted from further analysis. Such stations also have an unsatisfactory Coefficient of Determination for the regression (below $R^2 = 0.85$).

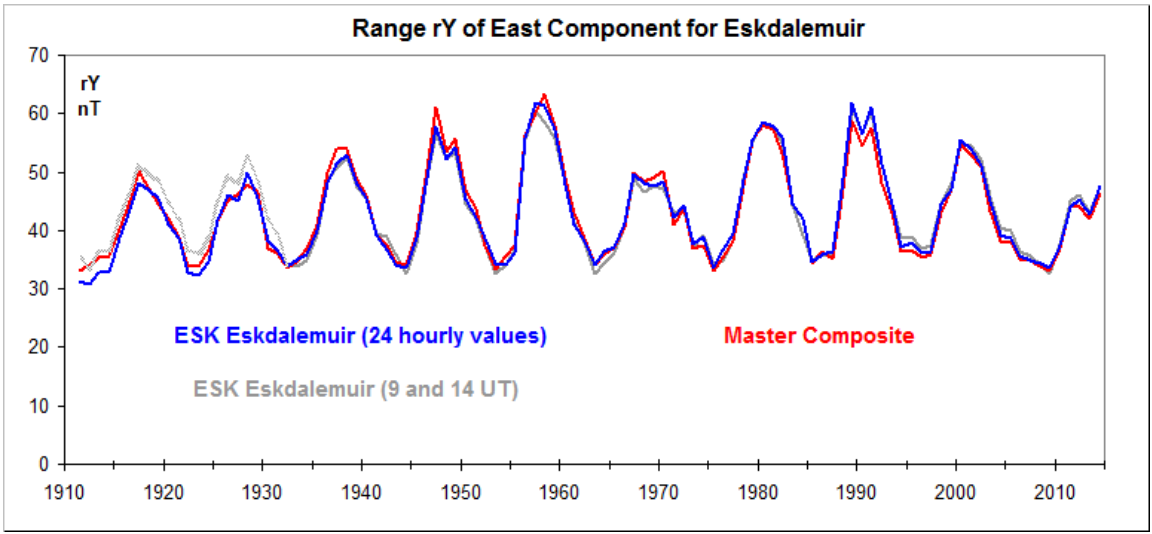

Figure 9: The unbroken series of *rY* ranges of the East–Component for Eskdalemuir (ESK, blue) scaled to the Master Composite (red), based on hourly averages 1911–2014 using the slope (1.055) from Figure 8. Using only the two values at $9^h$ and $14^h$ local time (happens to be UT) the grey curve results (the scaling factor is 8% higher).

If hourly averages or hourly spot–values are available, the range is calculated simply as the difference between the largest and the smallest hourly values of the yearly average curve. Because the curves are much alike (*c.f.* Figure 10) for stations between latitudes 15° and ~62°, varying mainly in amplitude, only two values at fixed hours during the day time are actually needed to determine the daily range, as shown in Figure 9 (the grey curve).

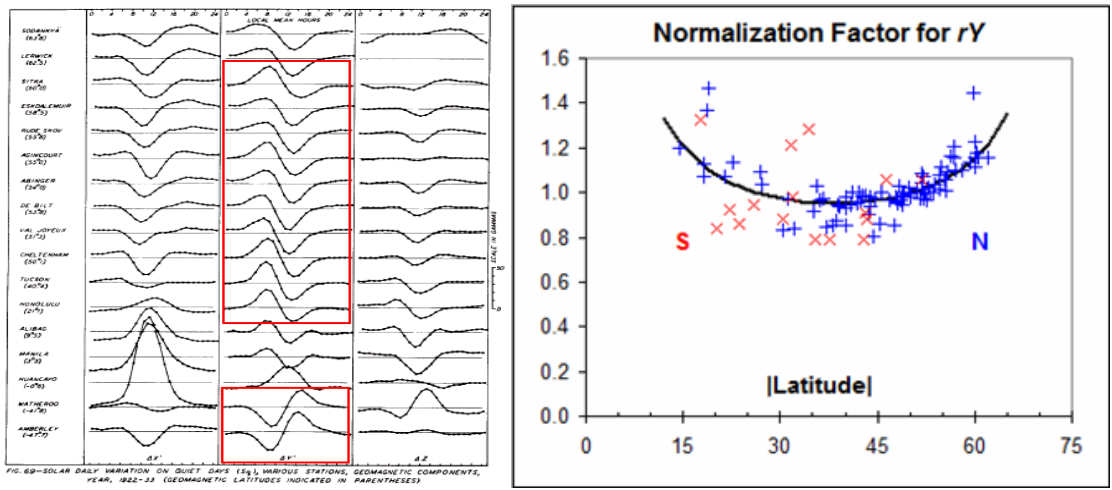

Figure 10: (Left) Diurnal variation of the three geomagnetic components, *X*, *Y* (middle), and *Z*, organized according to latitude Red boxes show how the variation is very similar for a broad range of latitudes. (Right) The normalization factor as a function of latitude (blue: Northern Hemisphere, red: Southern Hemisphere).

Many observatories operating during the 19$^{th}$ century were, in fact, only observing a few times per day. The fit to the Master Composite is nearly as close as for the full 24–hour coverage, because the two hours chosen are near the average times of maximum effect. There is small systematic discrepancy for ESK before 1932 due to a change of data reduction in 1932 (MacMillan and Clarke, 2011). Because such meta–information is rarely available, we generally make no attempt to correct for known and unknown minor changes. Large discrepancies or variance cause us simply to reject the data in question.

### 7.1. On 'Homogeneous' Data

There is an important, if somewhat philosophical, point to be made here about the misguided notion (Lockwood *et al.*, 2013) that using only a *single* station at any one time instead of all available and relevant data is somehow inherently 'better' because the single–station series would be more 'homogeneous'. This fallacy (leading to erroneous conclusions, *e.g.* Lockwood *et al.* (1999), Svalgaard *et al.* (2004), Clilverd *et al.* (2005), and Lockwood *et al.* (2013)) ignores the possibilities of (unreported, unknown, or disregarded) changes of observing procedure and data handling (MacMillan and Clarke, 2011), of changes of instrumental scale values (Svalgaard, 2014), and of influences of changing local conditions or aging instruments (Malin, 1996; Lockwood *et al.*, 2013). The situation is analogous to the clear superiority of the high–quality world–wide *am* index (Mayaud, 1967), constructed from a carefully calibrated global network of 24 stations, over the limited–quality *aa* index (Mayaud, 1972) constructed from only a single pair of stations. The *u* measure, if constructed from single–station data at a time (Lockwood *et al.*, 2013), the *aa* index constructed from a single station–pair at a time, and the Zürich sunspot number *Rz* constructed from observations largely at a single station (Waldmeier, 1971) were all optimistically, and somewhat pompously, claimed to be 'homogenous', but turned out not to be so, when re-examined critically and compared with multi–station or multi–index reconstructions (Svalgaard, 2014; Svalgaard and Cliver, 2007; Lockwood *et al.*, 2006; Lockwood *et al.*, 2014a; Clette *et al.*, 2014). Claims of homogeneity can only be made after extensive cross–checking with *other* datasets. We take the view, which as a bonus also allows an estimate of data uncertainties, that more data, carefully vetted, is better than less data.

### 7.2. A Secondary Master Record

As the Master Record only goes back to 1884 there is a need for a secondary master record going further back in order that we can normalize and utilize the earliest data (there is a vast amount of observational data (*Schering*, 1889) from the 19$^{th}$ century still awaiting digitization and analysis) that may not have overlapping coverage with the primary master record. So we continue this, somewhat tedious, section with a description of the construction of the secondary record. A number of stations (Prague PRA, Helsinki HLS (Nevanlinna, 2004), Milan MIL, Oslo OSL (Wasserfall, 1948), Colaba CLA, Vienna WIE, Munich MHN, Clausthal KLT, and St. Petersburg SPE) cover the interval before 1884 and also overlap with the master record. We can therefore normalize the records from those stations the usual way to the Master Composite and obtain by averaging the normalized records the sought after secondary master record, Figure 10, firmly connecting the two master records. Neither master record show any discernable trend of the sunspot cycle minimum values (dashed lines in Figures 7 and 11).

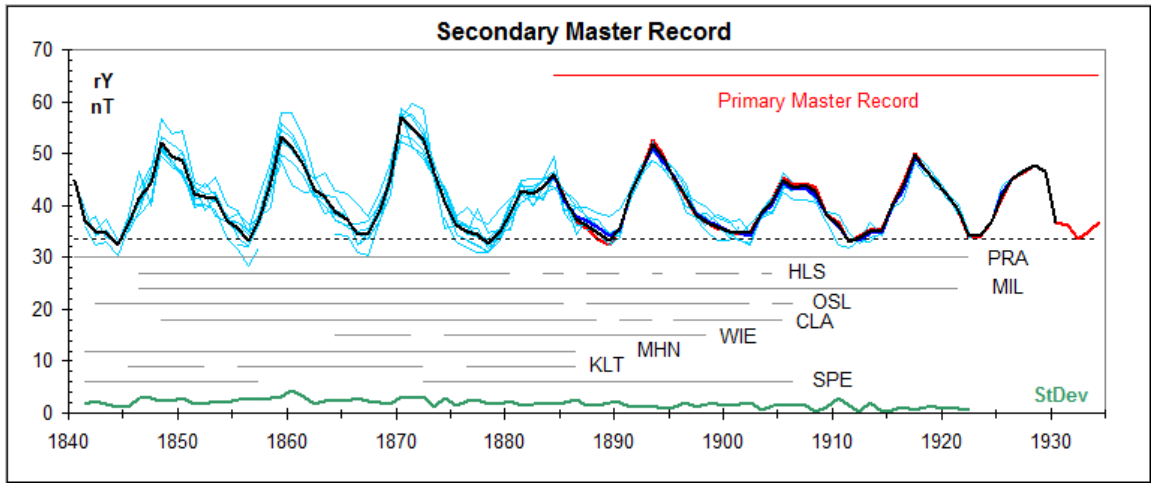

Figure 11: The stations PRA, HLS, MIL, OSL, CLA, WIE, MHN, KLT, and SPE have data that overlap with the Master Record (coverage shown by bars labeled with the station code) and we can thus normalize their records to the Master Composite (red curve). Normalized records for individual stations are shown with thin turquoise curves (average curve: blue). The thick black curve shows the final average when the overlapping part of the Master Composite is included, forming the secondary master record.

A source of unwanted variability is that metadata is often lacking as to which days and which hours were used to determine the ranges reported: all days, or only quiet days (and then which ones), what times of the day (including night hours if they suffered a substorm, creating a local extremum), and when, or if, such procedural details changed. We assume that the normalization absorbs enough of the effect of such changes that we can consider them to be akin to 'noise', whose average influence diminishes as the number of stations increases.

### 7.3. The Composite *rY* Record

Normalizing the ranges from the following 129 observatories to the Master Record(s) yields the composite shown in Figure 12.

```
AAA  ABG  ABN  AGN  AML  AMS  API  AQU  ARS  ASP  BAL  BDV  BEL
BER  BFE  BJI  BMT  BOU  BOX  BSL  CAO  CBI  CDP  CLA  CLF  CLH
CNB  COI  CTA  CTO  DBN  DOU  EBR  EKT  ELT  ESA  ESK  EYR  FRD
FRN  FUR  GCK  GEN  GLM  GNA  GRW  HAD  HBK  HBT  HER  HLP  HLS
HON  HRB  IRT  ISK  JAI  KAK  KDU  KEW  KLT  KNY  KNZ  KSH  LER
LNN  LOV  LRM  LVV  LZH  MAB  MBO  MIL  MIZ  MMB  MNH  MNK  MON
MZL  NCK  NEW  NGK  NUR  NVS  ODE  OSL  OTT  PAF  PAG  PET  PHI
PIL  POT  PRA  PSM  PST  QIX  ROM  RSV  SED  SFS  SIT  SJG  SPE
SSH  SVD  TAM  TFS  THJ  THY  TOK  TOO  TOR  TRW  TUC  UPS  VAL
VIC  VLJ  VQS  WAT  WHN  WIA  WIE  WIK  WIT  WLH  WNG  YAK
```

Table 3: IAGA 3-letter codes identifying observing station as listed in *e.g.* Rasson (2001).

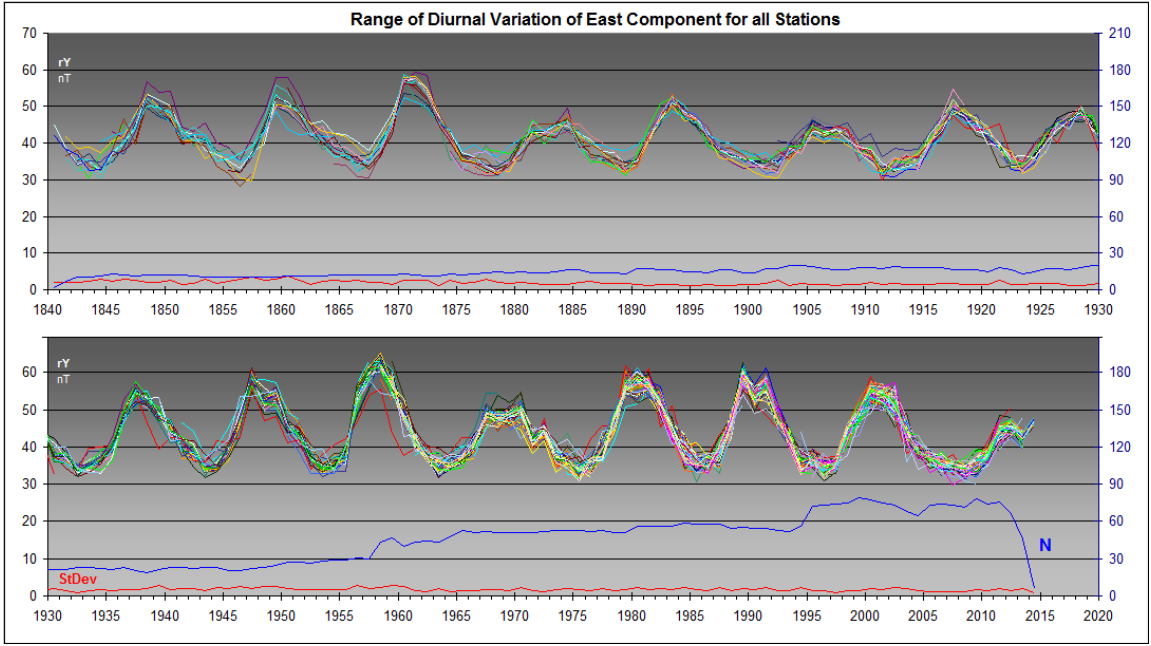

Figure 12: Normalized yearly average range, *rY* in nT, of the geomagnetic East component for each of the 129 stations used in the present paper. Different stations are plotted with different colors. The standard deviation is shown by the red curve at the bottom of each panel and the number of stations, *N*, for each station by the blue curve.

The normalization removes the dependence on latitude and most of the variation due to differences in underground conductivity. There remains the (minor) influence of geomagnetic activity in the auroral zone and the Ring Current as we did not limit ourselves to so-called 'quiet days'. The multi-colored Figure 12 shows that the 'spread' between stations is rather uniformly about four times the standard deviation, *SD*, corresponding to encompassing 95% of the data; this justifies computing the standard error *SE* of the mean as $SE = SD/\sqrt{N}$. Employing this device leads to Figure 13:

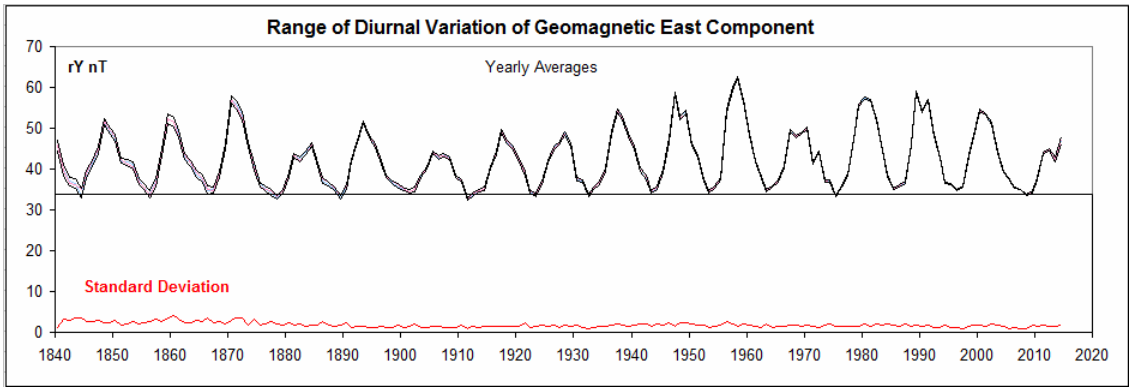

Figure 13: Normalized yearly average range of the diurnal variation, *rY* in nT, of the geomagnetic East component. The standard deviation is shown by the red curve at the bottom of the Figure and the standard ±1-σ error of the mean surrounds the average (red) curve. The annual values of *rY* are given in Table 1 at the end of the paper.

## 8. Relationship with Solar EUV and F10.7

We can now compare the observed composite range *rY* with the theoretical expectation that it be proportional to the square root of the EUV flux (and its proxy the F10.7 microwave flux), Figure 14:

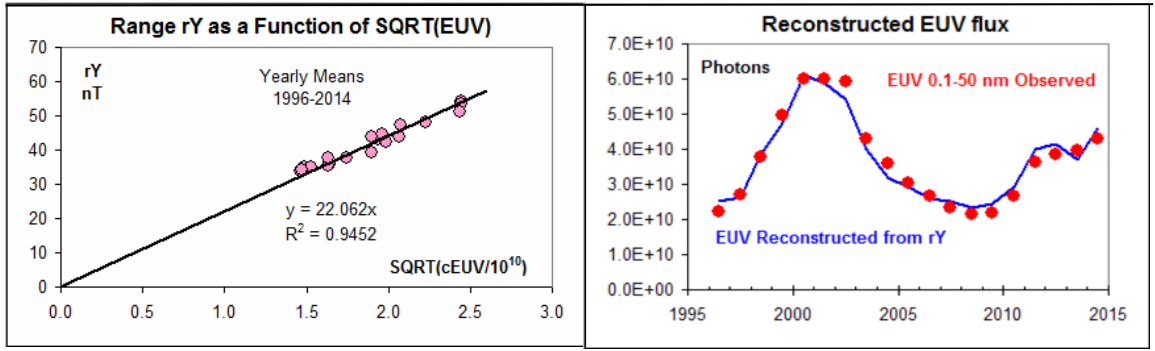

Figure 14: (Left) Yearly average ranges *rY* plotted against the square root of the corrected EUV flux for the years 1996-2014 (see section 3). The offset is negligible so there is simple proportionality as expected. (Right) EUV flux reconstructed from *rY* for 1996-2014 using the slope of the regression line.

The F10.7 microwave flux shows the same square root relationship with *rY* as already noted by Yamazaki and Kosch (2014). Since measurements of F10.7 go back to 1947 we can extend the regression plot that far back as well, Figure 15.

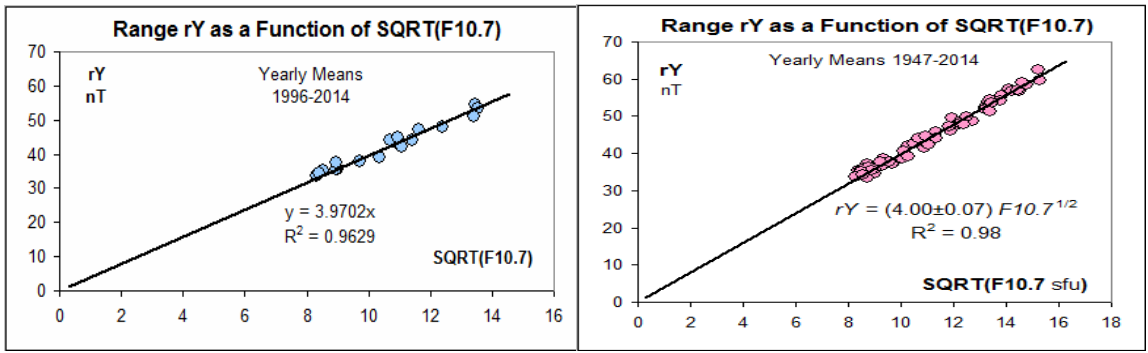

Figure 15: (Left) Yearly average ranges *rY* plotted against the square root of the F10.7 flux for the years 1996-2014 (*c.f.* section 4). The offset is negligible so there is simple proportionality as expected. (Right) Extending the regression back to the beginning of the F10.7 series in 1947.

At this point we have established the calibration factors for *rY* to reconstruct the EUV and F10.7 fluxes in their respective physical units. The tightness of the correlations and the nice homoskedacity (uniform variance) justify using the relationships in reverse and calculate the EUV and F10.7 fluxes from *rY*:

$$EUV = (rY/22)^2 \; 10^{10} \text{ photons (0.1-50 nm)}$$

$$F10.7 = (rY/4)^2 \text{ sfu}$$

Figure 16 shows how successful this procedure is. The reconstruction of F10.7 back to 1947 is excellent and justifies extending the reconstruction all the back to 1840.

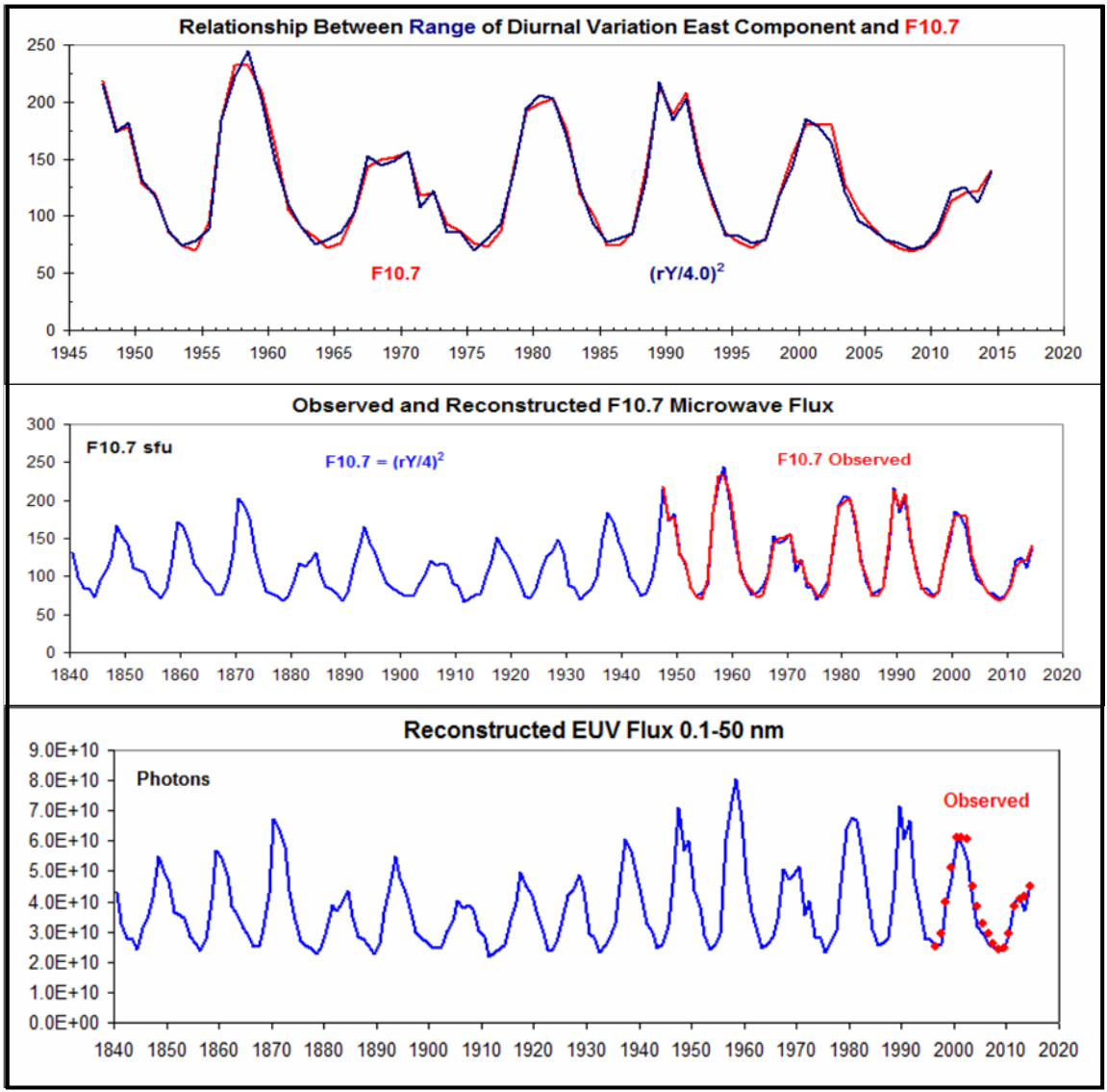

Figure 16: (Top) Yearly average values of the F10.7 flux (blue) compared to the reconstructed values (red) for 1947-2014. (Center) Same, but including the whole period 1840-2014. (Bottom) Yearly average values of the 0.1-50 nm reconstructed EUV flux (blue) and the observed flux (red dots).

The $2.5 \cdot 10^{10}$ photons/cm$^2$/sec EUV flux in the 0.1-50 nm wavelength range inferred for every sunspot minimum the past 175 years appears to be a 'basal' flux, present when visible solar activity has died away. The lack of any variation of this basal flux suggests that the flux (and the network causing it) is always there, presumably also during Grand Minima. If the magnetic network is always present, this means that a chromosphere is also a permanent feature, consistent with the observations of the 'red flash' observed during the 1706 solar eclipse (Young, 1881). This is, however, a highly contentious issue (Riley *et al.*, 2015), but one of fundamental importance.

As the magnetic field in the solar wind (the Heliosphere) ultimately arises from the magnetic field on the solar surface filtered through the corona, one would expect, at least an approximate, relationship between the network field and the Heliospheric field, the latter now firmly constrained (Svalgaard, 2015). Figure 17 shows a comparison of the *rY* proxy for the EUV flux from the surface network magnetic field structures, connected in the higher solar atmosphere to the coronal magnetic field, and then carried out into the Heliosphere to be observed near the Earth.

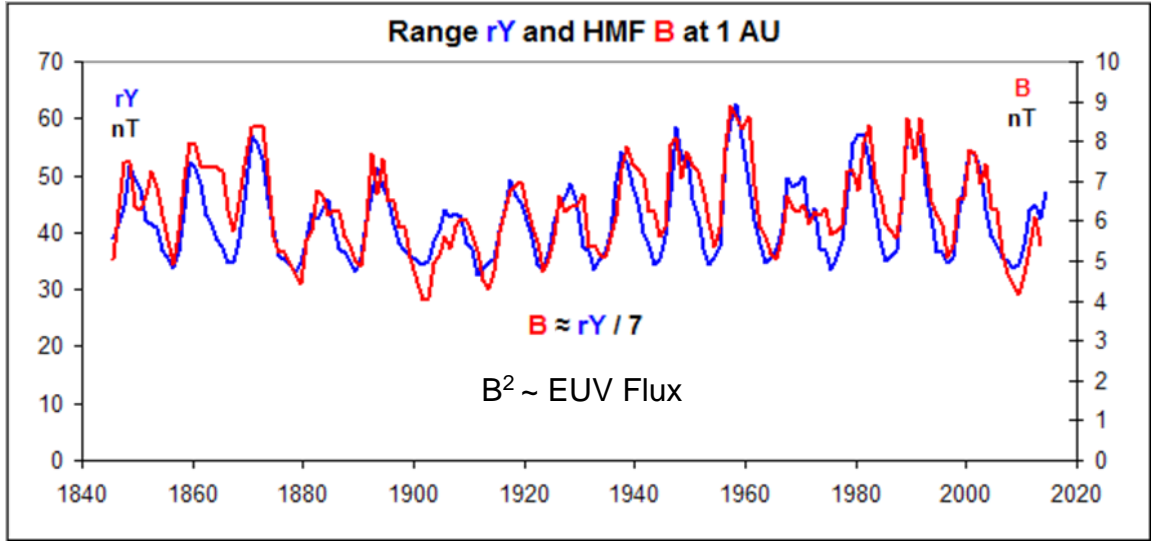

Figure 17: Yearly average values of the diurnal range *rY* of the geomagnetic East component (blue, left-hand scale) compared to the inferred magnitude of the Heliospheric magnetic field, *B*, near the Earth since the 1840s (red, right-hand scale).

Assuming that the EUV flux results from release of stored magnetic energy and therefore scales with the energy of the network magnetic field ($B^2$), we can understand the correspondence between the Heliospheric field and the network field. Again we are faced with the puzzle that there seems to be a 'floor' in both and with the question what happens to this floor during a Grand Minimum.

## 9. A Historical Interlude

When Rudolf Wolf discovered the relationship between his Relative Sunspot Number and the range of the diurnal variation of the Declination he at once realized that the relationship forded an independent check of the sunspot number and proceeded to collect and to request variation data from observers at (the often newly established) geomagnetic observatories and to compare the observations with his Relative numbers from year to year. At times the geomagnetic data would arrive belatedly and Wolf would *predict* from his relationship what the range would be and he was generally correct. In 1870 Wolf became 'alarmed' (Loomis, 1873) because the computed and observed variations seriously disagreed and Wolf, being so convinced that the relationship was real and physical and should be obeyed, consequently (Wolf, 1872) adjusted his method of comparing sunspot observers in order to make the anomaly go away such as to restore the agreement between the solar and the terrestrial data. Wolf continued to collect geomagnetic data until his death, and his successor, Wolfer, carried on until 1922 when finally the geomagnetic comparisons were discontinued as some participating observatories were shut down.

A factor that perhaps also contributed to the abandonment of the geomagnetic comparisons was that the relationship appeared to be changing with time such that the original coefficients were no longer applicable, thus undermining the rationale for comparing the solar and terrestrial data; the influence of the Sun seemed to be steadily diminishing, Figure 18 (top panel).

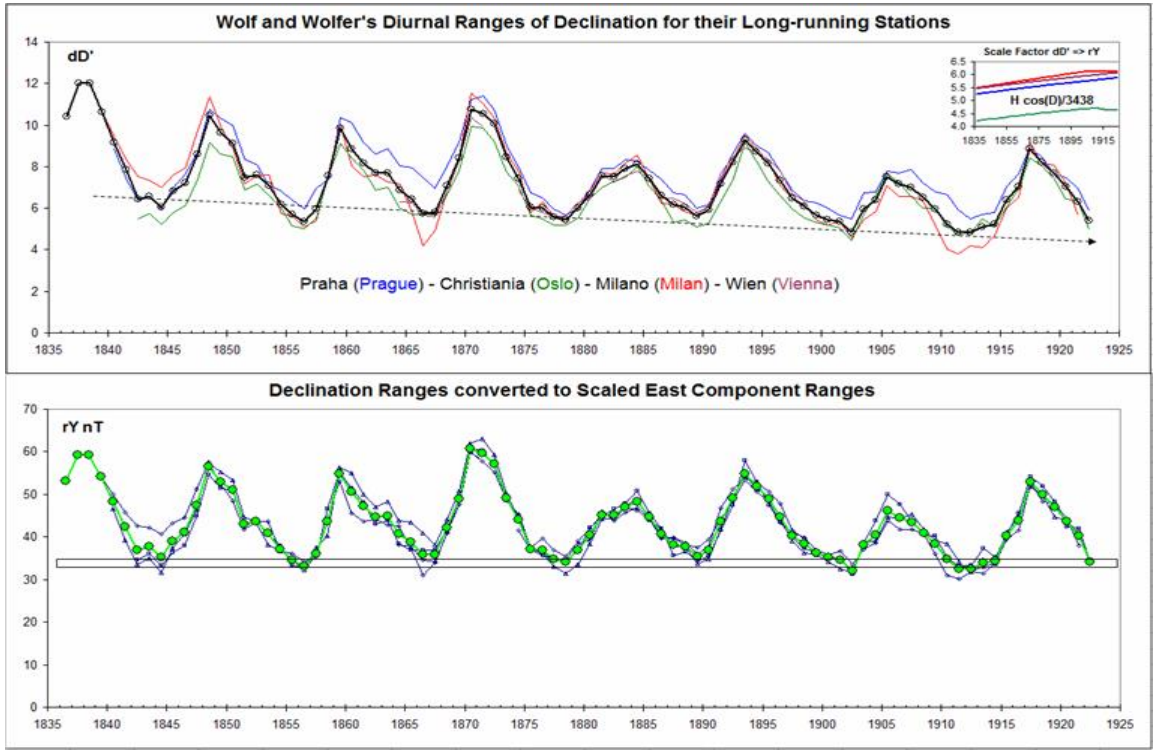

Figure 18: (Top) Diurnal range, *dD*, of Declination reported by Wolf and Wolfer for the four long-running stations: Prague, Oslo, Milan, and Vienna. The black curve with circle symbols shows the average of the four stations. The range shows a clear secular decrease, casting doubt on the physical meaning of the sunspot-range relationship. The inset shows the secular change of the conversion factors from *dD* to *rY* for the four stations, due to the changing main geomagnetic field. (Bottom) Taking the secular change of the conversion factors into account removes the secular change of the geomagnetic response, *rY*, and restores the relationship as well as reducing the spread from station to station (the average is shown by the green curve and symbols). Note, that the values at each solar minimum are very similar (horizontal bar).

Today we know that the relevant parameter for the geomagnetic response is the East Component, *Y*, rather than the Declination, *D*. Converting *D* to *Y* (using $Y = H \sin(D)$ and $rY = H \cos(D)\, dD$) restores the stable correlation without any significant long-term drift of the base values. So Rudolf Wolf was right, after all.

## 10. Earliest Observations of the Diurnal Range

Even before the 'Magnetic Crusade' of the 1840s we have scattered observations of the diurnal variation of the Declination. A detailed discussion of the early data will be the subject of a separate paper. Here we shall limit ourselves to early data mainly collected and published by Wolf and reduced by Loomis (1870, 1873) to the common scale of Prague, Figure 19.

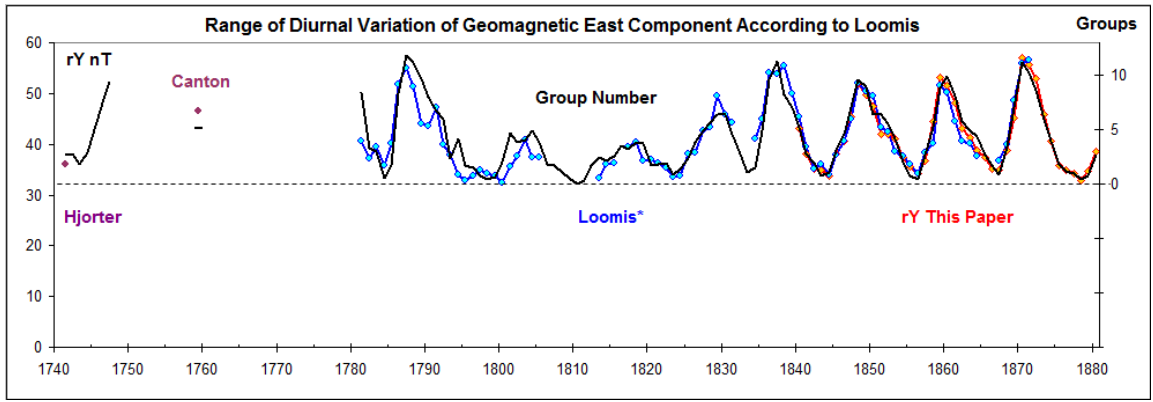

Figure 19: The range, $rY$, of the diurnal variation of the geomagnetic East component determined from the daily range of Declination (given by Loomis, 1870, 1873) converted to force units in the East direction (blue curve) and then scaled to match $rY$ (this paper, red curve), supplemented with observations by Canton (1759) and Hjorter (1747). The Sunspot Group Number (Svalgaard and Schatten, 2015) is shown (black curve without symbols) for comparison scaled (right-hand scale) to match $rY$.

Loomis drew two important and prescient conclusions: 1) the basal part of the "diurnal inequality (read: variation), amounting at Prague to six minutes is independent of the changes in the sun's surface from year to year", and 2) "the excess of the diurnal inequality above six minutes as observed at Prague, is almost exactly proportional to the amount of spotted surface upon the sun, and may therefore be inferred to be produced by this disturbance of the sun's surface, or both disturbances may be ascribed to a common cause". It is encouraging that the Sunspot Group Number series seems to agree well with diurnal range series, even for the earliest geomagnetic measurements. Loomis' conclusions are fully supported by our modern data and analyses.

Olof Peter Hjorter (with Anders Celsius) made ~10,000 observations of the diurnal variation of the Declination during 1740-1747 (Hjorter, 1747) at Uppsala, Sweden. Hjorter's (and Celsius') measurements were made with an instrument manufactured by Graham in London and the data are accurate to about one minute of arc and are the earliest data of sufficient quality and extent to allow firm determination of the diurnal variation. The original notebooks with observations have been preserved (and kindly made available to us by Olof Beckman, Uppsala) and a detailed analysis will be reported in a separate paper (Svalgaard and Beckman, 2015; a summary to be incorporated in the present one). At this point we only note that the variation at the sunspot minimum in 1741 was very similar to the variation at nearby Lovö in 1997, Figure 20.

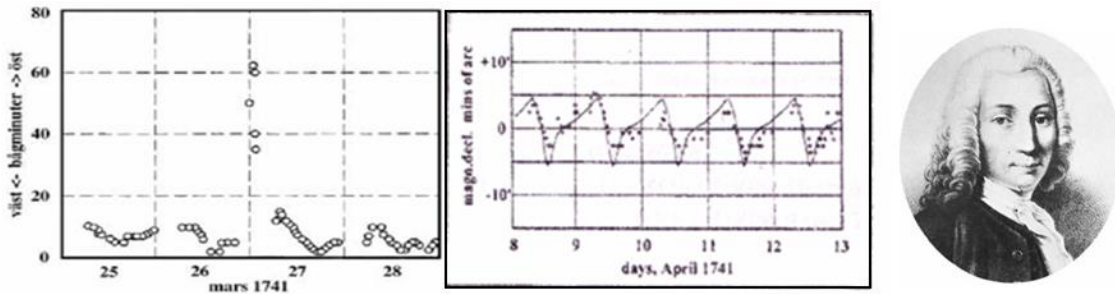

Figure 20: Observations by Hjorter of the variation of the Declination at Uppsala during the spring of 1741. The diurnal range was about 10 arc minutes, comparable to that at nearby Lovö magnetic observatory in April 1997 (full drawn curve in the middle panel). The observations of the large disturbance on March 27 (old style), 1741 were obtained during a great auroral display also observed by Graham in London, proving that auroral and magnetic phenomena were connected and were not just local effects.

## 11. Comparison with the Sunspot Group Series

Although it is important to stress that the Sunspot Group Number series (Svalgaard and Schatten, 2015) is a pure solar index and that the Diurnal Range series (Svalgaard, this paper) is a pure terrestrial index, it is also important to compare the two series to check for disagreements or differing trends. After all, we are concerned with quantifying manifestations of the long-time variation of the same underlying cause, the Sun's magnetic field. In order to compare the series we first put them on the same scale by regressing $rY$ against the group number $GN$, as shown in the right-hand panel of Figure 21. Then we can plot the series for easy visual comparison and also take the ratio for a numerical measure of the similarity, Figure 21.

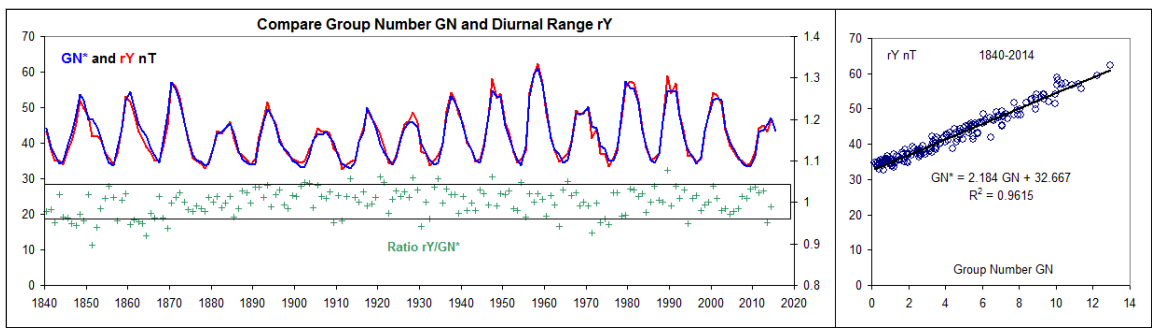

Figure 21: The Group Number (blue curve) scaled to match the Diurnal Range (red curve) using the regression equation obtained in the right-hand panel. The ratio (green symbols) between the two measures is 1.00±0.04 (box).

The ratio between the diurnal range and the scaled group number is slightly smaller than unity during 1840-1870, but is still within the combined error bar for the two series, so the geomagnetic data are excellent complements to the direct count of sunspot groups. Accepting this, justifies constructing a composite of $GN$ and $rY$ in terms of the group count so we can compare with the sunspot number, Figure 22. We consider this composite to be a 'true' representation of 'solar activity', or to alternatively *define* 'solar activity' because of how closely it correlates with the F10.7 microwave flux.

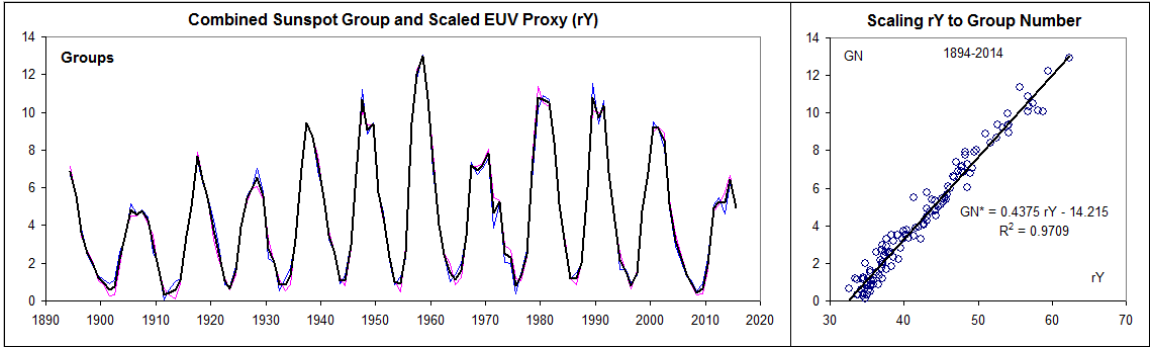

Figure 22: A composite Group Number (black curve) constructed as the average of the observed Group Number, *GN*, (pink curve) and the EUV proxy (*rY*, blue curve) scaled to *GN* according to the regression equation shown in the right-hand panel. We chose the time after Wolf's death in 1893 to exclude possible contamination or uncertainty from the use of his small telescope (*c.f.* Figure 4 of Svalgaard and Schatten (2015)).

The first step is to scale the composite Group Number series, *GN'*, to the Relative Sunspot Number, *SSN*. Hoyt and Schatten (1998) found the scaling factor to be given by *SSN* = 12.08 *GN*. By sheer coincidence we find the scaling factor to be 12.09 (right-hand panel of Figure 23) for the time interval 1894-1946. The reason for this choice is that there is good evidence that Max Waldmeier (1948) introduced a weighting of sunspots according to size and complexity in 1947 (section 5.2 of Clette et al. (2014), Svalgaard and Cagnotti (2015)). Figure 23 shows the observed SSN (blue curve) as reported by SILSO and the scaled values of GN (red curve; for convenience we drop the prime mark from now on).

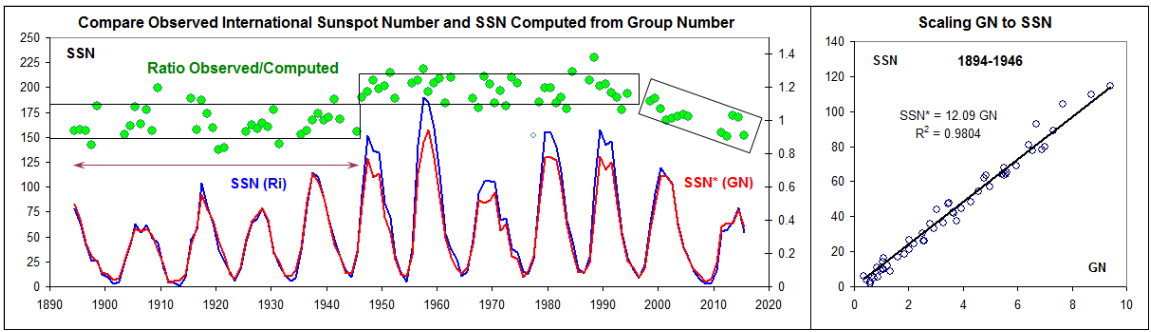

Figure 23: Comparison between the observed International Relative Sunspot Number series (blue curve, *SSN(Ri)*) and the composite Group Number series scaled (red curve, *SSN\*(GN)*) to match *SSN(Ri)* during the interval 1894-1946 (using the regression equation from the right-hand panel). For years where both series have values larger than 20 (to avoid the large noise resulting from ratios of small numbers) we plot the ratio between them (green dots with right-hand scale).

The large 'boxes' contain 90% of the data points. The average ratio for the box 1894-1946 is (not surprisingly) 0.990, but for the 1947-1994 box the average ratio is 1.192, which is an increase by a factor 1.204, giving a measure of the inflation of the reported SSN from 1947 onwards. This matches the average inflation derived from direct counting of spots, with and without weighting (section 5.2 of Clette *et al.* (2014); Svalgaard and Cagnotti (2015)).

What *is* notable, however, is the steady decline of the ratio from at least about 1995 to the present time. This discrepancy between the reported *SSN* and the 'true' solar activity (as measured by the equivalent F10.7, *rY*, and *GN* indices that all correlate so well with each other) has been noted before, *e.g.* by Svalgaard and Hudson (2010) and others. To get a better indication of the details of the decline we increase the time resolution from one year to one month. At the higher resolution the relationship between the *SSN* and F10.7 is no longer nearly linear. The right-hand panel of Figure 24 shows a $2^{nd}$-order fit to the data during the Zürich-era, assuming that the data are homogenous enough throughout that time. We shall use that fit to construct a synthetic *SSN* to compare with the reported *SSN*, Figure 24.

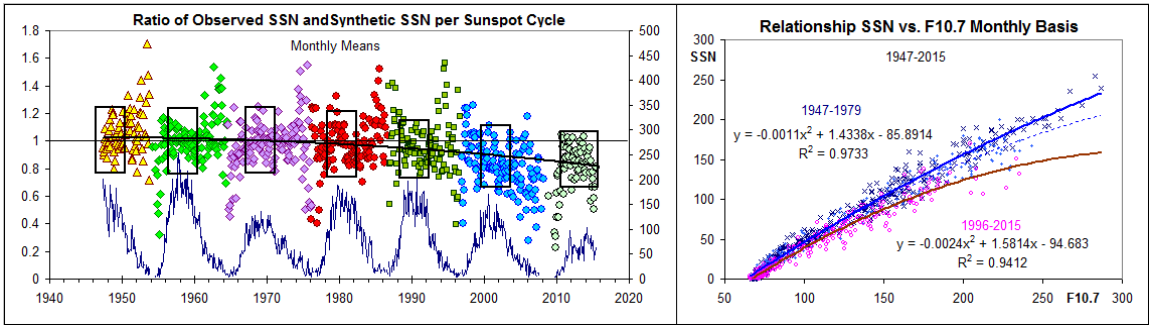

Figure 24: Monthly values color-coded per sunspot cycle of the ratio between the reported Sunspot Number (itself shown at the bottom of the Figure) and a Synthetic Sunspot Number derived from the F10.7 microwave flux using the regression equation for the interval 1947-1979 given in the right-hand panel (blue curve).

The monthly data show the same steady decline of the ratio so we'll have to accept that this is a real effect. What could be the cause of this decline of the Sunspot Number compared to F10.7? Has the weighting of sunspots been abandoned since ~1994? No, the analysis in Clette *et al.* (2014) shows that it has not. In addition, the observers in Locarno since August 2014 report both the weighted count and the un-weighted (actual) count of sunspots. Figure 25 shows the observed weight factor computed from their recent reports.

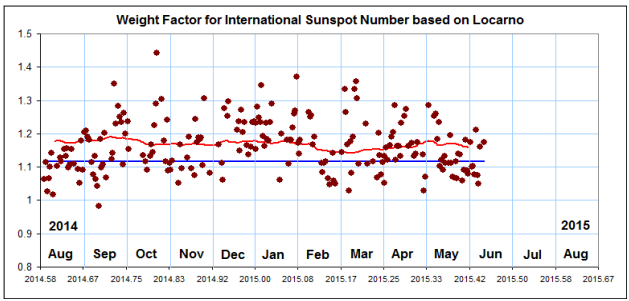

Figure 25: The ratio of the weighted *SSN* and the un-weighted (real) *SSN* reported by Locarno (brown dots) for each daily observation since August, 2014. The red curve shows the expected 27-day average weight factor computed from the formula in Clette *et al.* (2014). The blue line: see text.

For this low to medium level of sunspot activity (average *SSN* = 64) the average weight factor was 1.161, well above the suggested result (1.116, blue line) of the invalid analysis by Lockwood *et al.* (2014b), so we have to look elsewhere for an explanation of the decline.

### 11.1. Number of Spots per Group is Not Constant

The basic idea behind the Group Sunspot Number was that the number of spots per group is constant. Even in Wolf's definition of the Relative Sunspot Number = 10 Groups + Spots that assumption is built-in, as the factor of 10 for the groups is held constant. We can investigate the validity of this assumption for recent solar cycles using data from the German SONNE network of sunspot observers (Bulling, 2013) and Swiss reference station. As we know, to this day the Locarno observers weight larger spots stronger than small spots, so the weighted spot count will on average be 30-50% larger than the raw count where each spot is counted only once as in Wolf's and Wolfer's original scheme (Wolfer, 1907: "Notiert ein Beobachter mit seinem Instrumente an irgend einem Tage *g* Fleckengruppen mit insgesamt *f* Einzelflecken, ohne Rücksicht auf deren Grösse, so ist die daraus abgeleitete Relativzahl jenes Tages $r = k(10g+f)$"). The SONNE observers do not employ weighting: each spot is counted only once. It is important that for both groups of observers, the counting methods (albeit different) have been unchanged over the period of interest.

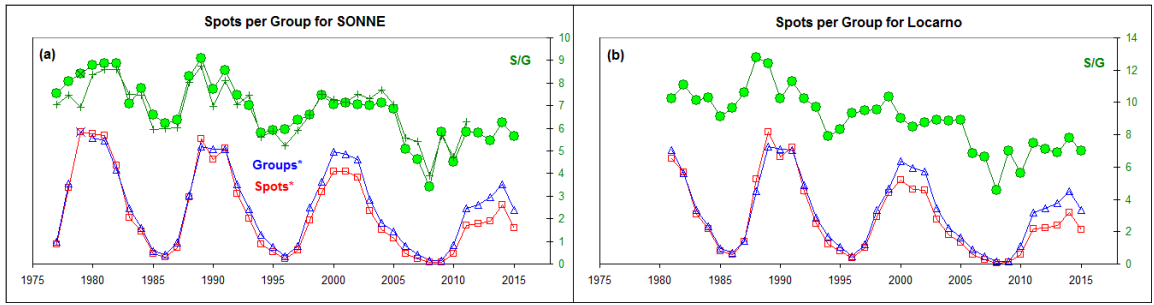

Figure 26: The number of spots per group as a function of time (green dots) for the ~500,000 individual observations made by the SONNE network (left) and for Locarno (right). The green curve with pluses shows the ratio derived from the raw counts, not normalized with *k*-factors, and yet not significantly different. The lower part of the panels shows the variation of number of groups (blue triangles) and of the number of spots (red squares) both scaled to match each other before 1992. Note for both series the decreasing spot count, relative to the group count.

Figure 26 shows that the average number of spots per group has been decreasing steadily for both SONNE and Locarno and is therefore not due to drifts of calibration or decreasing visual acuity of the primary Locarno observer (Sergio Cortesi). If the 'missing spots' were large spots with significant magnetic flux one would expect F10.7 and *rY* to decrease as well, contrary to the observed trends (Figures 23-24), so the missing spots must be the smallest spots, as also suggested by Lefèvre and Clette (2011). It appears that this may be a natural explanation for the decline of the Sunspot Number compared to F10.7 and *rY*.

## 11.2. Comparison with Other UV proxies

The emission core of the Magnesium II doublet (λ = 280 nm) exhibits the largest natural solar irradiance variability above 240 nm. The Mg II doublet is a broad absorption feature with narrow emission peaks in the core. Radiation in the line wings originates in the photosphere and shows much less variability. Therefore, the ratio of line core intensity to wing intensity provides a good estimate of solar variability because the use of an intensity ratio cancels degradation effects. The core-to-wing ratio is frequently used as a proxy for spectral solar irradiance variability from the UV to EUV. The so-called 'Bremen' composite series covering 1978-2015 (Snow *et al.*, 2014) utilizes all available satellite data, Figure 27.

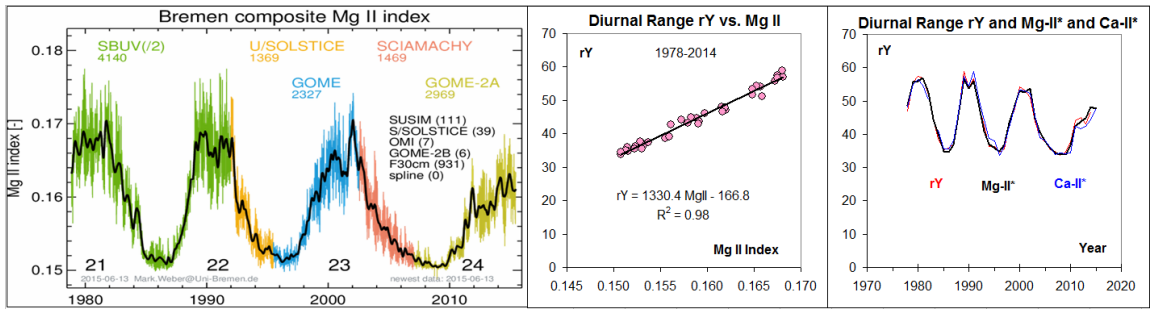

Figure 27: (Left) The Bremen Mg II Index composite (courtesy Mark Weber, with permission). (Middle) The yearly averages of the Bremen index have a very high correlation ($R^2 = 0.98$) with our *rY* composite. (Right) Scaling the Mg II index and the (bit more noisy) NSO Ca II K-line index (λ = 393 nm) to the Diurnal Range, *rY*, shows that all three indices agree well over the range from EUV to low-λ visible.

As the Relative Sunspot Number as currently defined deviates from the EUV-UV measures it is no longer the usual faithful representation of solar activity. Whether that was also the case at times in the past, *e.g.* during Grand Minima, is an open and intriguing question. The Group Number, on the other hand, tracks the UV indices closely and appears to be a good proxy for solar surface magnetic fields, at least for the past two and a half centuries, and again it is not clear what happens during a Grand Minimum.

## 12. Conclusions

The Diurnal Range, *rY*, of the geomagnetic East component can be determined with confidence from observatory data back to 1840 and estimated with reasonable accuracy a century further back in time. The range *rY* correlates very strongly with the F10.7 microwave flux and with a range of measures of the EUV-UV flux and thus with the solar magnetic field giving rise to these manifestations of solar activity. The variation of the range also matches closely that of the Sunspot Group Number and the Heliospheric magnetic field, but is at variance with the usual Relative Sunspot Number for the past two solar cycles, which we ascribe to a progressive deficit of small sunspots, such that the number of spots per group is not constant, but has been steadily decreasing. The range (and thus the magnetic activity causing it) reaches a constant (non-zero) floor at every solar minimum for which we have data.

Table 1: Yearly values of the Diurnal Range in nT of the Geomagnetic East Component with low and high 1-σ limits reflecting the standard error of the mean. Also listed are the reconstructed values of the F10.7 flux.

| Year | Low | Mean | High | F10.7 | Year | Low | Mean | High | F10.7 |
|---|---|---|---|---|---|---|---|---|---|
| 1840.5 | 42.11 | **43.15** | 44.19 | 116.4 | 1928.5 | 48.30 | **48.61** | 48.92 | 147.7 |
| 1841.5 | 37.38 | **38.05** | 38.73 | 90.5 | 1929.5 | 45.43 | **45.76** | 46.10 | 130.9 |
| 1842.5 | 34.41 | **35.04** | 35.67 | 76.8 | 1930.5 | 37.13 | **37.60** | 38.07 | 88.4 |
| 1843.5 | 34.24 | **34.96** | 35.68 | 76.4 | 1931.5 | 36.82 | **37.13** | 37.44 | 86.2 |
| 1844.5 | 33.01 | **33.85** | 34.69 | 71.6 | 1932.5 | 33.51 | **33.71** | 33.91 | 71.0 |
| 1845.5 | 37.40 | **38.01** | 38.62 | 90.3 | 1933.5 | 34.82 | **35.14** | 35.47 | 77.2 |
| 1846.5 | 39.78 | **40.50** | 41.22 | 102.5 | 1934.5 | 36.13 | **36.52** | 36.92 | 83.4 |
| 1847.5 | 44.74 | **45.40** | 46.06 | 128.8 | 1935.5 | 39.54 | **39.86** | 40.19 | 99.3 |
| 1848.5 | 51.55 | **52.07** | 52.60 | 169.5 | 1936.5 | 48.70 | **49.04** | 49.38 | 150.3 |
| 1849.5 | 48.99 | **49.83** | 50.67 | 155.2 | 1937.5 | 53.71 | **54.13** | 54.54 | 183.1 |
| 1850.5 | 46.62 | **47.55** | 48.48 | 141.3 | 1938.5 | 52.02 | **52.48** | 52.94 | 172.1 |
| 1851.5 | 41.27 | **41.92** | 42.57 | 109.8 | 1939.5 | 47.65 | **48.25** | 48.85 | 145.5 |
| 1852.5 | 41.46 | **42.05** | 42.64 | 110.5 | 1940.5 | 44.78 | **45.17** | 45.55 | 127.5 |
| 1853.5 | 40.11 | **41.10** | 42.09 | 105.6 | 1941.5 | 39.49 | **39.94** | 40.39 | 99.7 |
| 1854.5 | 36.78 | **37.34** | 37.91 | 87.2 | 1942.5 | 37.66 | **38.11** | 38.57 | 90.8 |
| 1855.5 | 34.89 | **35.69** | 36.50 | 79.6 | 1943.5 | 34.28 | **34.61** | 34.94 | 74.9 |
| 1856.5 | 33.26 | **34.11** | 34.97 | 72.7 | 1944.5 | 35.01 | **35.50** | 36.00 | 78.8 |
| 1857.5 | 35.77 | **36.79** | 37.81 | 84.6 | 1945.5 | 39.14 | **39.57** | 40.00 | 97.9 |
| 1858.5 | 43.34 | **44.40** | 45.46 | 123.2 | 1946.5 | 46.79 | **47.40** | 48.02 | 140.4 |
| 1859.5 | 51.98 | **53.14** | 54.30 | 176.5 | 1947.5 | 57.74 | **58.15** | 58.57 | 211.4 |
| 1860.5 | 50.18 | **51.46** | 52.73 | 165.5 | 1948.5 | 52.15 | **52.67** | 53.19 | 173.4 |
| 1861.5 | 47.35 | **48.21** | 49.06 | 145.2 | 1949.5 | 53.57 | **54.07** | 54.56 | 182.7 |
| 1862.5 | 42.70 | **43.13** | 43.56 | 116.3 | 1950.5 | 45.46 | **45.83** | 46.20 | 131.3 |
| 1863.5 | 40.62 | **41.25** | 41.87 | 106.3 | 1951.5 | 42.78 | **43.14** | 43.50 | 116.3 |
| 1864.5 | 38.10 | **38.84** | 39.58 | 94.3 | 1952.5 | 37.06 | **37.42** | 37.77 | 87.5 |
| 1865.5 | 36.76 | **37.47** | 38.18 | 87.7 | 1953.5 | 34.16 | **34.48** | 34.79 | 74.3 |
| 1866.5 | 34.35 | **35.10** | 35.85 | 77.0 | 1954.5 | 35.12 | **35.42** | 35.73 | 78.4 |
| 1867.5 | 34.36 | **34.92** | 35.47 | 76.2 | 1955.5 | 37.29 | **37.62** | 37.95 | 88.5 |
| 1868.5 | 38.17 | **38.71** | 39.25 | 93.6 | 1956.5 | 53.70 | **54.21** | 54.72 | 183.6 |
| 1869.5 | 44.58 | **45.11** | 45.64 | 127.2 | 1957.5 | 59.04 | **59.43** | 59.82 | 220.8 |
| 1870.5 | 56.36 | **57.06** | 57.75 | 203.5 | 1958.5 | 61.96 | **62.29** | 62.63 | 242.5 |
| 1871.5 | 54.99 | **55.73** | 56.48 | 194.1 | 1959.5 | 56.35 | **56.77** | 57.19 | 201.4 |
| 1872.5 | 52.09 | **52.88** | 53.67 | 174.8 | 1960.5 | 47.89 | **48.29** | 48.69 | 145.8 |
| 1873.5 | 45.35 | **45.81** | 46.26 | 131.1 | 1961.5 | 41.74 | **41.96** | 42.19 | 110.1 |
| 1874.5 | 40.00 | **40.70** | 41.39 | 103.5 | 1962.5 | 37.78 | **37.95** | 38.12 | 90.0 |
| 1875.5 | 35.42 | **35.88** | 36.34 | 80.5 | 1963.5 | 34.50 | **34.79** | 35.09 | 75.7 |
| 1876.5 | 34.42 | **34.96** | 35.49 | 76.4 | 1964.5 | 35.48 | **35.66** | 35.83 | 79.5 |
| 1877.5 | 33.58 | **34.30** | 35.02 | 73.5 | 1965.5 | 36.86 | **37.05** | 37.25 | 85.8 |
| 1878.5 | 32.37 | **32.88** | 33.39 | 67.6 | 1966.5 | 40.42 | **40.61** | 40.81 | 103.1 |
| 1879.5 | 34.22 | **34.71** | 35.19 | 75.3 | 1967.5 | 48.97 | **49.23** | 49.48 | 151.5 |
| 1880.5 | 38.11 | **38.58** | 39.05 | 93.0 | 1968.5 | 47.64 | **47.87** | 48.10 | 143.2 |
| 1881.5 | 42.89 | **43.35** | 43.81 | 117.5 | 1969.5 | 48.41 | **48.61** | 48.82 | 147.7 |
| 1882.5 | 42.14 | **42.54** | 42.94 | 113.1 | | 1970.5 | 49.52 | **49.83** | 50.14 | 155.2 |
| 1883.5 | 43.76 | **44.18** | 44.59 | 122.0 | | 1971.5 | 41.13 | **41.32** | 41.52 | 106.7 |
| 1884.5 | 45.64 | **46.02** | 46.40 | 132.3 | | 1972.5 | 44.10 | **44.24** | 44.38 | 122.3 |
| 1885.5 | 40.82 | **41.32** | 41.81 | 106.7 | | 1973.5 | 36.93 | **37.16** | 37.39 | 86.3 |
| 1886.5 | 36.75 | **37.38** | 38.01 | 87.3 | | 1974.5 | 36.70 | **36.97** | 37.25 | 85.4 |
| 1887.5 | 36.21 | **36.62** | 37.04 | 83.8 | | 1975.5 | 33.17 | **33.41** | 33.66 | 69.8 |
| 1888.5 | 34.76 | **35.17** | 35.57 | 77.3 | | 1976.5 | 35.59 | **35.81** | 36.02 | 80.1 |
| 1889.5 | 33.48 | **33.95** | 34.42 | 72.0 | | 1977.5 | 38.34 | **38.61** | 38.89 | 93.2 |
| 1890.5 | 35.07 | **35.43** | 35.80 | 78.5 | | 1978.5 | 46.89 | **47.10** | 47.31 | 138.6 |
| 1891.5 | 41.67 | **41.94** | 42.21 | 109.9 | | 1979.5 | 55.49 | **55.73** | 55.97 | 194.1 |
| 1892.5 | 46.39 | **46.72** | 47.05 | 136.4 | | 1980.5 | 57.08 | **57.38** | 57.68 | 205.8 |
| 1893.5 | 51.21 | **51.59** | 51.96 | 166.3 | | 1981.5 | 56.66 | **56.88** | 57.09 | 202.2 |
| 1894.5 | 47.35 | **47.73** | 48.12 | 142.4 | | 1982.5 | 51.45 | **51.74** | 52.02 | 167.3 |
| 1895.5 | 45.19 | **45.50** | 45.80 | 129.4 | | 1983.5 | 44.30 | **44.55** | 44.79 | 124.0 |
| 1896.5 | 41.00 | **41.38** | 41.76 | 107.0 | | 1984.5 | 38.32 | **38.62** | 38.91 | 93.2 |
| 1897.5 | 37.88 | **38.15** | 38.42 | 91.0 | | 1985.5 | 35.01 | **35.25** | 35.49 | 77.7 |
| 1898.5 | 36.52 | **36.82** | 37.11 | 84.7 | | 1986.5 | 35.72 | **35.92** | 36.12 | 80.6 |
| 1899.5 | 35.25 | **35.59** | 35.92 | 79.2 | | 1987.5 | 36.76 | **37.06** | 37.35 | 85.8 |
| 1900.5 | 34.65 | **35.03** | 35.40 | 76.7 | | 1988.5 | 45.82 | **46.02** | 46.23 | 132.4 |
| 1901.5 | 34.19 | **34.56** | 34.93 | 74.7 | | 1989.5 | 58.57 | **58.85** | 59.12 | 216.4 |
| 1902.5 | 34.45 | **35.03** | 35.62 | 76.7 | | 1990.5 | 53.75 | **53.98** | 54.21 | 182.1 |
| 1903.5 | 38.05 | **38.34** | 38.62 | 91.8 | | 1991.5 | 56.43 | **56.74** | 57.04 | 201.2 |
| 1904.5 | 40.06 | **40.39** | 40.73 | 102.0 | | 1992.5 | 47.87 | **48.07** | 48.26 | 144.4 |
| 1905.5 | 43.96 | **44.31** | 44.66 | 122.7 | | 1993.5 | 42.49 | **42.68** | 42.86 | 113.8 |
| 1906.5 | 42.72 | **43.08** | 43.44 | 116.0 | | 1994.5 | 36.04 | **36.33** | 36.62 | 82.5 |
| 1907.5 | 43.20 | **43.49** | 43.79 | 118.2 | | 1995.5 | 36.15 | **36.31** | 36.47 | 82.4 |
| 1908.5 | 42.33 | **42.70** | 43.07 | 114.0 | | 1996.5 | 34.56 | **34.73** | 34.90 | 75.4 |
| 1909.5 | 38.03 | **38.36** | 38.70 | 92.0 | | 1997.5 | 35.39 | **35.51** | 35.62 | 78.8 |
| 1910.5 | 36.49 | **36.99** | 37.50 | 85.5 | | 1998.5 | 42.94 | **43.11** | 43.28 | 116.1 |
| 1911.5 | 32.19 | **32.56** | 32.92 | 66.2 | | 1999.5 | 47.62 | **47.80** | 47.97 | 142.8 |
| 1912.5 | 33.40 | **33.82** | 34.24 | 71.5 | | 2000.5 | 54.01 | **54.22** | 54.43 | 183.7 |
| 1913.5 | 34.48 | **34.84** | 35.19 | 75.8 | | 2001.5 | 53.10 | **53.29** | 53.48 | 177.5 |
| 1914.5 | 34.78 | **35.15** | 35.53 | 77.2 | | 2002.5 | 50.77 | **51.06** | 51.34 | 162.9 |
| 1915.5 | 39.95 | **40.30** | 40.66 | 101.5 | | 2003.5 | 43.49 | **43.74** | 44.00 | 119.6 |
| 1916.5 | 43.88 | **44.30** | 44.72 | 122.7 | | 2004.5 | 38.98 | **39.17** | 39.36 | 95.9 |
| 1917.5 | 49.03 | **49.50** | 49.98 | 153.2 | | 2005.5 | 37.39 | **37.52** | 37.65 | 88.0 |
| 1918.5 | 46.46 | **46.85** | 47.24 | 137.2 | | 2006.5 | 35.31 | **35.45** | 35.59 | 78.5 |
| 1919.5 | 44.74 | **45.14** | 45.55 | 127.4 | | 2007.5 | 34.77 | **34.90** | 35.04 | 76.1 |
| 1920.5 | 41.88 | **42.24** | 42.61 | 111.5 | | 2008.5 | 33.55 | **33.68** | 33.81 | 70.9 |
| 1921.5 | 38.62 | **39.26** | 39.90 | 96.3 | | 2009.5 | 34.41 | **34.60** | 34.80 | 74.8 |
| 1922.5 | 34.08 | **34.47** | 34.85 | 74.2 | | 2010.5 | 37.46 | **37.62** | 37.77 | 88.4 |
| 1923.5 | 33.79 | **34.18** | 34.57 | 73.0 | | 2011.5 | 43.95 | **44.16** | 44.37 | 121.9 |
| 1924.5 | 36.14 | **36.63** | 37.13 | 83.9 | | 2012.5 | 44.82 | **45.01** | 45.20 | 126.6 |
| 1925.5 | 41.25 | **41.67** | 42.09 | 108.5 | | 2013.5 | 42.79 | **43.08** | 43.36 | 116.0 |
| 1926.5 | 44.88 | **45.28** | 45.68 | 128.1 | | 2014.5 | 46.29 | **46.69** | 47.09 | 136.3 |
| 1927.5 | 45.92 | **46.19** | 46.45 | 133.3 | | 2015.5 | | | | |

**Acknowledgements**

We acknowledge the use of data from the following sources: 1) CELIAS/SEM experiment on the Solar Heliospheric Observatory (SOHO) spacecraft, a joint European Space Agency (ESA), United States National Aeronautics and Space Administration (NASA) mission. 2) The Laboratory for Atmospheric and Space Physics (CU) TIMED Mission. 3) The Solar Radio Monitoring Programme at Dominion Radio Astrophysical Observatory operated jointly by National Research Council, Canada and Natural Resources, Canada. 4) The Nobeyama Radio Observatory, NAOJ, Japan. 5) World Data Centers for Geomagnetism in Kyoto and Edinburgh. 6) Data collected at geomagnetic observatories by national institutes according to the high standards of magnetic observatory practice promoted by INTERMAGNET (www.intermagnet.org). 7) Data collected by Wolf and Wolfer in *Mittheilungen*. 8) Yearbooks from the British Geological Survey http://www.geomag.bgs.ac.uk/data_service/data/yearbooks/yearbooks.html. 9) World Data Center for the production, preservation and dissemination of the international sunspot number http://sidc.be/silso/. 10) *Wasserfall* (1948). 11) The SONNE Network http://sonne.vdsastro.de/index.php?page=gem/res/results.html#provrel. 12) The Bremen composite Mg II index http://www.iup.uni-bremen.de/gome/gomemgii.html.

We have benefited from comments by Ingrid Cnossen and Ed Cliver. We thank Vladimir Papitashvilli for the CORRGEOM program to compute the geomagnetic field elements for the years 1590–1995. This research has made use of NASA's Astrophysics Data System. LS thanks Stanford University for support.

**References**

Allen, C.W.: 1948, Critical frequencies, sunspots, and the Sun's ultra-violet radiation, *Terr. Magn. Atmos. Electr.* **53**(4), 433–448, doi:10.1029/TE053i004p00433

Appelton, E.W.: 1947, http://www.nobelprize.org/nobel_prizes/physics/laureates/1947/appleton-lecture.pdf

Bulling, A.: 2013, The SONNE Sunspot Number Network – 35 Years & Counting, *3rd SSN Workshop*, http://www.leif.org/research/SSN/Bulling.pdf

Canton, J.: 1759, An Attempt to Account for the Regular Diurnal Variation of the Horizontal Magnetic Needle; And Also for Its Irregular Variation at the Time of an Aurora Borealis, *Phil. Trans.* **51**, 398–445, doi:10.1098/rstl.1759.0040

Chapman, S., Gupta, J.C., Malin, S.R.C.: 1971, The Sunspot Cycle Influence on the Solar and Lunar Daily Geomagnetic Variations, *Proc. Roy. Soc. Lond.* **324**(1566), 1–15, doi:10.1098/rspa.1971.0124

Chree, C.: 1913, Some Phenomena of Sunspots and of Terrestrial Magnetism at Kew Observatory, *Phil. Trans. Roy. Soc. Lond. A* **212**, 75–116, doi: 10.1098/rsta.1913.0003

Clette, F., Svalgaard, L., Vaquero, J.M., Cliver, E.W.: 2014, Revisiting the Sunspot Number – A 400–Year Perspective on the Solar Cycle, *Space Sci. Rev.* **186**, 35–103, doi:10.1007/s11214-014-0074-2

Clilverd, M.A., Clark, T.D.G., Clarke, E., Rishbeth, H.: 1998, Increased magnetic storm activity from 1868 to 1995, *J. Atmos. Solar–Terr. Phys.* **60**, 1047–1056, doi:10.1016/S1364-6826(98)00049-2

Clilverd, M.A., Clarke, E., Ulich, T., Linthe, J., Rishbeth, H.: 2005, Reconstructing the long-term aa index, *J. Geophys. Res.* **110**, A07205, doi:10.1029/2004JA010762

Cnossen, I., Richmond, A.D., Wiltberger, M.: 2012, The dependence of the coupled magnetosphere–ionosphere–thermosphere system on the Earth's magnetic dipole moment, *J. Geophys. Res.* **117**, A05302, doi:10.1029/2012JA017555

Didkovsky, L., Wieman, S.: 2014, Ionospheric total electron contents (TECs) as indicators of solar EUV changes during the last two solar minima, *J. Geophys. Res.* **119**(A), 1–10, doi:10.1002/2014JA019977

Dudok de Wit, T., Bruinsma, S., Shibasaki, K.: 2014, Synoptic radio observations as proxies for upper atmosphere modelling. *J. Space Weather Space Clim.* **4**, A06, doi:10.1051/swsc/2014003

Emmert, J.T., McDonald, S.E., Drob, D.P., Meier, R.R., Lean, J.L., Picone, J.M.: 2014, Attribution of interminima changes in the global thermosphere and ionosphere, *J. Geophys. Res.* **119**(A), doi:10.1002/2013JA019484

Fouassier, D., Chulliat, A.: 2009, Extending backwards to 1883 the French magnetic hourly data series, in *Proceedings of the XIIIth IAGA Workshop on Geomagnetic Observatory Instruments, Data Acquisition, and Processing*, U.S. Geological Survey Open-File Report 2009–**1226**, 86–94, J. J. Love, ed.

Gautier, J-A,: 1852, Notice sur quelques recherches récentes, astronomiques et physiques, relative aux apparences que présente le corps du solei, *Bibliothèque Universelle de Genève, Archives des sciences physiques et naturelles* **20**, 177–207, Ferd. Ramboz et Comp., Genève; http://tinyurl.com/mgs7hqw

Graham, G.: 1724, An Account of Observations Made of the Variation of the Horizontal Needle at London, in the Latter Part of the Year 1722, and Beginning of 1723, *Phil. Trans.* **33**, 96–107, doi:10.1098/rstl.1724.0020

Heaviside, O.: 1902, Telegraphy I Theory, *Encyclopedia Britannica (10 th ed.)* **33**, 213–218

Hjorter, O. P.: 1747, Om Magnet-Nålens åtskillige ändringar etc, *Kong. Svensk. Vet. Handl.* **8**, 27

Hoyt, D.V, Schatten, K.H.: 1998, Group sunspot numbers: a new solar activity reconstruction. *Solar. Phys.* **181**, 491–512

Ieda, A., Oyama, S., Vanhamäki, H., Fujii, R., Nakamizo, A., Amm, O., Hori, T., Takeda, M., Ueno, G., Yoshikawa, A., Redmon, R.J., Denig, W.F, Kamide, Y., Nishitani, N.: 2014, Approximate forms of daytime ionospheric conductance, *J. Geophys. Res. Space Physics* **119**, 10397–10415, doi:10.1002/2014JA020665

Jackson, A., Jonkers, A.R.T., Walker, M.R.: 2000, Four centuries of geomagnetic secular variation from historical records, *Phil. Trans. Roy. Soc. Lond.*, *A* **358**, 957-990,

doi:10.1098/rsta.2000.0569. PC–DOS program at
http://www.leif.org/research/CORRGEOM.EXE

Judge, D.L., McMullin, D.R., Ogawa, H.S., Hovestadt, D., Klecker, B., Hilchenbach, M., Möbius, E., Canfield, L.R., Vest, R.E., Watts, R., Tarrio, C., Kühne, M., Wurz, P.: 1998, First solar EUV irradiances obtained from SOHO by the CELIAS/SEM, *Solar Phys.* **177**, 161–173, doi:10.1023/A:1004929011427

Kennelly, A.E.: 1902, On the Elevation of the Electrically–Conducting Strata of the Earth's Atmosphere, *Elec. World & Eng.*, **39** 473–473

Koyama, Y., Shinbori. A., Tanaka, Y., Hori, T., Nosé, M., Oimatsu, S.: 2014, An Interactive Data Language software package to calculate ionospheric conductivity by using numerical models, *Computer Phys. Comm.* **185**, 3398–3405, doi:10.1016/j.cpc.2014.08.011

Lamont, J.v.: 1851, Ueber die zehnjährige Periode, welche sich in der Größe der täglichen Bewegung der Magnetnadel darstellt, *Ann. der Physik* **160**(12), 572–584, doi:10.1002/andp.18511601206

Lean, J.L., Warren, H.P., Mariska, J.T., Bishop, J.: 2003, A new model of solar EUV irradiance variability, 2, Comparisons with empirical models and observations and implications for space weather, *J. Geophys. Res.* **108**(A2), 1059, doi:10.1029/2001JA009238

Lean, J.L., Emmert, J.T., Picone, J.M., Meier, R.R.: 2011, Global and regional trends in ionospheric total electron content, *J. Geophys. Res.* **116**, A00H04, doi:10.1029/2010JA016378

Lefèvre, L., Clette, F.: 2011, A global small sunspot deficit at the base of the index anomalies of solar cycle 23, *Astr. & Astroph.* **536**, id.L11, 4 pp., doi:10.1051/0004-6361/201118034

Lockwood, M., Stamper, R., Wild, M.N.: 1999, A doubling of the sun's coronal magnetic field during the last 100 years, *Nature* **399**, 437–439, doi:10.1038/20867

Lockwood, M., Whiter, D., Hancock, B., Henwood, R., Ulich, T., Linthe, H.J., Clarke, E., Clilverd, M.: 2006, The long-term drift in geomagnetic activity: calibration of the aa index using data from a variety of magnetometer stations, *Rutherford Appleton Laboratory (RAL) Harwell Oxford, UK,* available at: http://bit.ly/KjeIio

Lockwood, M., Barnard, L., Nevanlinna, H., Owens, M.J., Harrison, R.G., Rouillard, A.P., Davis, C.J.: 2013, Reconstruction of geomagnetic activity and near–Earth interplanetary conditions over the past 167 yr – Part 1: A new geomagnetic data composite, *Ann. Geophys.* **31**(11), 1957–1977, doi:10.5194/angeo-31-1957-2013

Lockwood, M., Nevanlinna, H., Vokhmyanin, M, Ponyavin, D., Sokolov, S., Barnard, L., Owens, M.J., Harrison, R.G., Rouillard, A.P., Scott, C.J.: 2014a, Reconstruction of geomagnetic activity and near–Earth interplanetary conditions over the past 167 yr – Part 3: Improved representation of solar cycle 11, *Ann. Geophys.* **32**(4), 367–381, doi:10.5194/angeo-32-367-2014

Lockwood, M., Owens, M.J., Barnard, L.: 2014b, Centennial variations in sunspot number, open solar flux, and streamer belt width: 1. Correction of the sunspot number record since 1874, *J. Geophys. Res. Space Physics* **119**, 5172–5182, doi:10.1002/2014JA019970

Loomis, E.: 1870, Comparison of the mean daily range of Magnetic Declination, with the number of Auroras observed each year, and the extent of the black Spots on the surface of the Sun, *Am. Journ. Sci. Arts, 2$^{nd}$ Series* **50**(149), 153–171

Loomis, E.: 1873, Comparison of the mean daily range of the Magnetic Declination and the number of Auroras observed each year, *Am. Journ. Sci. Arts, 3$^{rd}$ Series* **5**(28), 245–260

Love, J.J., Rigler, E.J.: 2014, The magnetic tides of Honolulu, *Geophys. J. Int.* **197**(3), 1335–1353, doi:10.1093/gji/ggu090

Maeda, K.: 1977, Conductivity and drift in the ionosphere, *J. Atmos. Terr. Phys.* **39**, 1041–1053, doi:10.1016/0021-9169(77)90013-7

MacMillan, S., Droujinina, A.: 2007, Long-term trends in geomagnetic daily variation, *Earth Planets and Space* **59**, 391–395, doi:10.1186/BF03352699

MacMillan, S., Clarke, E.: 2011, Resolving issues concerning Eskdalemuir geomagnetic hourly values, *Ann. Geophys.* **29**, 283–288, doi:10.5194/angeo-29-283-2011

Malin, S.R.C.: 1973, Worldwide Distribution of Geomagnetic Tides, *Phil. Trans. Roy. Soc. Lond.*, *A* **274**, 551–594, doi:10.1098/rsta.1973.0076

Malin, S.R.C.: 1996, Geomagnetism at the Royal Observatory, Greenwich, *Q. J. Roy. Astr. Soc.* **37**, 65–74

Martini, D., Mursula, K., Orispää, M., Linthe, H. –J.: 2015, Long-term decrease in the response of midlatitude stations to high-speed solar wind streams in 1914–2000, *J. Geophys. Res. Space Physics* **120**, 2662–2674, doi: 10.1002/2014JA020813

Mayaud, P.–N.: 1965, Analyse morphologique de la variabilité jour–à–jour de la variation journalière "régulière" $S_R$ du champ magnétique terrestre, II – Le système de courants *CM* (Régions non–polaires), *Ann. de Géophys.* **21**, 514–544

Mayaud, P. –N.: 1967, Calcul préliminaire d'indices Km, Kn, Ks, ou Am, An, et As, mesures de l'activité magnétique à l'échelle mondiale et dans les hémisphères Nord et Sud, *Ann. de Géophys.* **23**(4), 585-617

Mayaud, P.–N.: 1972, The aa indices: A 100-year series characterizing the magnetic activity, *J. Geophys. Res.* **77**(34), 6870-6874, doi:10.1029/JA077i034p06870

Nevanlinna, H.: 2004, Results of the Helsinki magnetic observatory 1844–1912, *Ann. Geophys.* **22**(5), 1691–1704, doi:10.5194/angeo-22-1691-2004

Nusinov, A.A. (2006), Ionosphere as a natural detector for investigations of solar EUV flux variations, *Adv. Space Res.* **37**(2), 426- 432, doi:10.1016/j.asr.2005.12.001

Olsen, N.: 1996, A new tool for determining ionospheric currents from magnetic satellite data, *Geophys. Res. Lett.* **23**(24), 3635–3638, doi:10.1029/96GL02896

Rasson, J.L.: 2001, The status of the world-wide network of magnetic observatories, their location and instrumentation. *Contributions to Geophysics and Geodesy* **31**, 427–439

Richmond, A.D.: 1995, Ionospheric electrodynamics, in *Handbook of Atmospheric Electrodynamics* vol. **II**, edited by H. Volland, 249–290, CRC Press, Boca Raton, FL, ISBN:978-0849325205

Riley, P., Lionello, R., Linker, J.A., Cliver, E., Balogh, A., Beer, J., Charbonneau, P., Crooker, N., deRosa, M., Lockwood, M., Owens, M., McCracken, K., Usoskin, I., Koutchmy, S.: Inferring the Structure of the Solar Corona and Inner Hemisphere During the Maunder Minimum Using Global Thermodynamic Magnetohydrodynamic Simulations, *Ap. J.* **802**, 105, doi:10.1088/0004-637X/8-2/2/105

Sabine, E.: 1852, On Periodical Laws Discoverable in the Mean Effects of the Larger Magnetic Disturbances – No. II, *Phil. Trans. Roy. Soc. Lond.* **142**, 103–124, doi:10.1098/rstl.1852.0009

Samson, J.A.R., Gardner, J.L.: 1975, On the Ionization Potential of Molecular Oxygen, *Canadian J. Phys.* **53**(19), 1948–1952, doi:10.1139/p75-244

Schering, K.: 1889, Die Entwicklung und der gegenwartige Standpunkt der erdmagnetische Forschung, *Geograph. Jahrbuch* **13**, 171–220, http://www.leif.org/research/Schering-1889.pdf

Schwabe, S.H.: 1844, Sonnenbeobachtungen im Jahre 1843, *Astron. Nachricht.* **21**(495), 233–236

Schuster, A.: 1908, The Diurnal Variation of Terrestrial Magnetism, *Phil. Trans. Roy. Soc. London*, *A* **208**, 163–204, doi: 10.1098/rsta.1908.0017

Shibasaki, K., Ishiguro, M., Enome, S.: 1979, Solar Radio Data Acquisition and Communication System (SORDACS) of Toyokawa Observatory, *Proc. of the Res. Inst. of Atmospherics, Nagoya Univ.* **26**, 117–127

Snow, M., Weber, M., Machol, J., Viereck, R., Richard, E.: 2014, Comparison of Magnesium II core-to-wing ratio observations during solar minimum 23/24, *J. Space Weather Space Clim.* **4**, A04, doi: 10.1051/swsc/2014001

Stewart, B.: 1882, Hypothetical Views Regarding the Connexion between the State of the Sun and Terrestrial Magnetism, *Encyclopedia Britannica (9$^{th}$ ed.)* **16**, 181–184

Svalgaard, L., Cliver, E.W.: 2007, Interhourly variability index of geomagnetic activity and its use in deriving the long–term variation of solar wind speed, *J. Geophys. Res.* **112**(A10), doi:10.1029/2007JA012437

Svalgaard, L.: 2010, Sixty+ Years of Solar Microwave Flux, *SHINE Conference 2010*, Santa Fe, NM, http://www.leif.org/research/SHINE-2010-Microwave-Flux.pdf

Svalgaard, L.: 2010, Updating the Historical Sunspot Record, in *SOHO–23: Understanding a Peculiar Solar Minimum*, *ASP Conference Series* **428**, 297–305, S. R. Cranmer, J. T. Hoeksema, and J. L. Kohl, eds., Astronomical Society of the Pacific, San Francisco, CA, ISBN:978-1-58381-736-0

Svalgaard, L.: 2012, How well do we know the sunspot number? in *Comparative Magnetic Minima: Characterizing Quiet Times in the Sun and Stars*, *Proc. IAU Symposium* **286***,* 27–33, C. H. Mandrini and D. F. Webb, eds., doi:10.1017/S1743921312004590

Svalgaard, L.: 2014, Correction of errors in scale values for magnetic elements for Helsinki, *Ann. Geophys.* **32**, 633–641, doi:10.5194/angeo-32-633-2014

Svalgaard, L.: 2015, Reconstruction of Heliospheric Magnetic Field 1835-2015, *Solar Phys.* (submitted, this issue)

Svalgaard, L., Beckman, O.: 2015, Analysis of Hjorter's Observations 1740-1747 of Diurnal Range of Declination, *Solar Phys.* (submitted, this issue)

Svalgaard, L., Cagnotti, M.: 2015, The Effect of Weighting of Sunspot Counts, *Solar Phys.* (submitted, this issue)

Svalgaard, L., Cliver, E.W., Le Sager, P.: 2004, IHV: A new geomagnetic index, *Adv. Space Res.* **34**(2), 436–439

Svalgaard, L., Hudson, H.S.: 2010, The Solar Microwave Flux and the Sunspot Number, in *SOHO–23: Understanding a Peculiar Solar Minimum*, *ASP Conference Series* **428**, 325–328, S. R. Cranmer, J. T. Hoeksema, and J. L. Kohl, eds., Astronomical Society of the Pacific, San Francisco, CA, ISBN:978-1-58381-736-0

Svalgaard, L., Schatten, K.H.: 2015, Reconstruction of the Sunspot Group Number: the Backbone Method, *Solar Phys.* (submitted, this issue)

Tapping, K.F.: 1987, Recent solar radio astronomy at centimeter wavelengths: The temporal variability of the 10.7–cm flux, *J. Geophys. Res.* **92**(D1), 829–838, doi:10.1029/JD092iD01p00829

Tapping, K.F.: 2013, The 10.7 cm solar radio flux ($F_{10.7}$), *Space Weather* **11**, 394–406, doi:10.1002/swe.20064

Takeda, M.: 1991, Role of Hall conductivity in the ionospheric dynamo, *J. Geophys. Res.* **96**(A6), 9755–9759, doi:10.1029/91JA00667

Takeda, M.: 2013, Contribution of wind, conductivity, and geomagnetic main field to the variation in the geomagnetic Sq field, *J. Geophys. Res. Space Physics* **118**, 4516–4522, doi:10.1002/jgra.50386

Waldmeier, M.: 1948, 100 Jahre Sonnenfleckenstatistik, *Astron. Mittl. Eidg. Sternwarte. Zürich* **16**, No. 152, 1–8

Waldmeier, M.: 1971, An Objective Calibration of the Scale of Sunspot-Numbers, *Astron. Mitt. Eidg. Sternwarte Zürich* No **304**, 1–10

Wasserfall, K.F.: 1948, Discussion of data for magnetic declination at Oslo, 1843–1930, and before 1843, *Terr. Mag. Atmos. Electr.* **53**(3), 279–290, doi:10.1029/TE053i003p00279

Wieman, S.R., Didkovsky, L.V., Judge, D.L.: 2014, Resolving Differences in Absolute Irradiance Measurements Between the SOHO/CELIAS/SEM and the SDO/EVE, *Solar Phys.* **289**, 2907–2925, doi:10.1007/s11207-014-0519-5

Wolf, J.R.: 1852a, Entdeckung des Zusammenhanges zwischen den Declinationsvariationen der Magnetnadel und den Sonnenflecken, *Mitth. der naturforsch. Gesell. Bern 224–264* Nr. **245**, 179–184

Wolf, J.R.: 1852b, Vergleichung der Sonnenfleckenperiode mit der Periode der magnetische Variationen, *Mitth. der naturforsch. Gesell. Bern 224–264* Nr. **255**, 249–270

Wolf, J.R.: 1857, Beitrag zur Geschichte der Entdeckung des Zusammenhanges zwischen Erdmagnetismus und Sonnenflecken, *Mitth. über die Sonnenflecken* **III**, 27–50

Wolf, J.R.: 1859, Über die Möglichkeit aus den Sonnenflecken–Relativzahlen die erdmagnetische Declinationsvariationen vorauszuberechnen, *Mitth. über die Sonnenflecken* **IX**, 207–246

Wolf, J.R.: 1872, Beobachtungen der Sonnenflecken im Jahre 1871, sowie Berechnung der Relativzahlen und Variationen dieses und Neu-Berechnung derjenigen des vorhergehenden Jahres, *Mitth. über die Sonnenflecken* **XXX**, 381–414

Wolfer, A.: 1907, Die Haufigkeit und heliographische Verteilung der Sonnenflecken im Jahre 1906, *Astronomische Mitteilungen* **XCVIII**, 252–414

Woods, T.N., Eparvier, F.G., Bailey, S.M., Chamberlin, P.C., Lean, J.L., Rottman, G.J., Solomon, S.C., Tobiska, W.K., Woodraska, D.L.: 2005, The Solar EUV Experiment (SEE): Mission overview and first results, *J. Geophys. Res.* **110**, A01312, doi:10.1029/2004JA010765

Yamazaki, Y., Kosch, M.J.: 2014, Geomagnetic lunar and solar daily variations during the last 100 years, *J. Geophys. Res.* **119***A*, 1–10, doi:10.1002/2014JA020203

Young, C.A.: 1881, *The Sun*, D. Appleton, New York, 321pp, 182